\documentclass[twocolumn,twocolappendix]{aastex63}

\usepackage{amsmath}
\usepackage{soul}
\usepackage{ulem}
\definecolor{dgreen}{RGB}{26,148,49}
\definecolor{forest}{RGB}{3, 148, 49}

\usepackage{lineno}

\shorttitle{Is the Radio Source Dipole from NVSS Consistent with the CMB and $\Lambda$CDM?}
\shortauthors{Cheng, Chang \& Lidz}
\graphicspath{{./}{figures/}}

\begin{document}

\title{Is the Radio Source Dipole from NVSS Consistent with the CMB and $\Lambda$CDM?}

\author[0000-0002-5437-0504]{Yun-Ting Cheng}
\address{Jet Propulsion Laboratory, California Institute of Technology, 4800 Oak Grove Drive, Pasadena, CA 91109, USA}
\address{California Institute of Technology, 1200 E. California Boulevard, Pasadena, CA 91125, USA}

\author[0000-0001-5929-4187]{Tzu-Ching Chang}
\address{Jet Propulsion Laboratory, California Institute of Technology, 4800 Oak Grove Drive, Pasadena, CA 91109, USA}
\address{California Institute of Technology, 1200 E. California Boulevard, Pasadena, CA 91125, USA}

\author[0000-0002-3950-9598]{Adam Lidz}
\affiliation{Center for Particle Cosmology, Department of Physics and Astronomy, University of Pennsylvania, Philadelphia, PA 19104, USA}
\address{Jet Propulsion Laboratory, California Institute of Technology, 4800 Oak Grove Drive, Pasadena, CA 91109, USA}

\begin{abstract}
The dipole moment in the angular distribution of the cosmic microwave background (CMB) is thought to originate from the Doppler effect and our motion relative to the CMB frame. Observations of large-scale structure (LSS) should show a related ``kinematic dipole'' and help test the kinematic origin of the CMB dipole. Intriguingly, many previous LSS dipole studies suggest discrepancies with the expectations from the CMB. Here we reassess the apparent inconsistency between the CMB measurements and dipole estimates from the NVSS catalog of radio sources. We find that it is important to account for the shot noise and clustering of the NVSS sources, as well as kinematic contributions, in determining the expected dipole signal. We use the clustering redshift method and a cross-matching technique to refine estimates of the clustering term. We then derive a probability distribution for the expected NVSS dipole in a standard $\Lambda$CDM cosmological model including all (i.e., kinematic, shot-noise and clustering) dipole components. Our model agrees with most of the previous NVSS dipole measurements in the literature at better than $\lesssim 2\sigma$. We conclude that the NVSS dipole is consistent with a kinematic origin for the CMB dipole within $\Lambda$CDM.
\end{abstract}


\section{Introduction} \label{S:intro}
The Cosmological Principle, one of the fundamental assumptions of the standard $\Lambda$CDM cosmological model, states that the universe is homogeneous and isotropic when averaged over sufficiently large scales. However, the cosmic microwave background (CMB) exhibits a dipole anisotropy at a level of $\Delta T/T\sim10^{-3}$, which is two orders of magnitude larger than the anisotropies on smaller scales ($\Delta T/T\sim10^{-5}$). This dipole is commonly interpreted as owing to our motion with respect to the CMB rest frame.

A crucial independent test of this kinematic hypothesis is to measure the dipole moment in the angular distribution of a large-scale structure (LSS) tracer at lower redshift. The overall mass distribution and the LSS are expected to share the same rest frame as the CMB, and hence the LSS should contain a matching dipole anisotropy from our motion with respect to the CMB/LSS frame. Hereafter, we refer to this as the ``kinematic LSS dipole'' to distinguish it from other contributions, as will be discussed shortly, to the LSS dipole moments. An inconsistency between the CMB dipole and the kinematic LSS dipole would be intriguing, and provide a potential additional handle on the physics of the early universe, with the possible resolutions depending on the nature of any discrepancy. For example, one possibility is that the CMB dipole may be partly intrinsic as opposed to entirely of kinematic origin. This requires superhorizon isocurvature perturbations, however, since adiabatic modes do not produce a CMB dipole anisotropy \citep{1990ApJ...364..341P,1991PhRvD..44.3737T,1996PhRvD..53.2908L,2008PhRvD..78h3012E,2014PhRvD..89f3518G}. Furthermore, a viable scenario here would need to respect the stringent bounds on isocurvature modes from higher-$\ell$ Planck CMB measurements \citep{2020A&A...641A..10P,2020A&A...641A...9P}. Also, note that the smaller-scale (higher-$\ell$) CMB anisotropies are modulated and aberrated by our motion with respect to the CMB frame. These signatures have been detected by the Planck collaboration at $4-\sigma$ significance, and the results are fully consistent with a kinematic origin for the CMB dipole, although there is still room for other possibilities  \citep{2014A&A...571A..27P}. For example, \citet{2021JCAP...07..035D} argue that a superhorizon adiabatic fluctuation can enhance the LSS dipole, while preserving the CMB dipole. On the contrary, \citet{2022JCAP...10..019D} find that a superhorizon adiabatic fluctuation doe not impact the LSS dipole. Instead, they suggest that a larger LSS dipole requires an enhanced local peculiar velocity and a compensating superhorizon isocurvature perturbation to avoid overproducing the CMB dipole. Detecting or constraining such possibilities, among others, provide strong motivations for LSS dipole measurements.

Towards this end, there is a long history of using radio source catalogs to probe the LSS dipole signal. \citet{1984MNRAS.206..377E} first proposed measuring the LSS dipole using radio source counts, and the first such detection was achieved by \citet{2002Natur.416..150B} using the National Radio Astronomy Observatory (NRAO) Very Large Array (VLA) Sky Survey \citep[NVSS; ][]{1998AJ....115.1693C}. While this early measurement found consistency with the prediction based on the CMB dipole, subsequent analyses using the same NVSS dataset as well as other radio surveys have reported significantly higher dipole amplitudes than expected \citep[e.g., ][]{2011ApJ...742L..23S,2012MNRAS.427.1994G,2013AA...555A.117R,2015APh....61....1T,2016JCAP...03..062T,2017MNRAS.471.1045C,2018JCAP...04..031B,2021ApJ...908L..51S,2022ApJ...937L..31S,2023arXiv230305141S,2023arXiv230515335W}. One exception is the recent study by \citet{2022ApJ...931L..14D} which used radio source catalogs from the Very Large Array Sky Survey \citep[VLASS; ][]{2020PASP..132c5001L} and the Rapid Australian Square Kilometer Array Pathfinder Continuum Survey \citep[RACS; ][]{2020PASA...37...48M} across $90\%$ of the sky, finding consistency with the expected CMB dipole. However, subsequent investigations by \citet{2022ApJ...937L..31S} and \citet{2023MNRAS.524.3636S} analyzed the same two datasets separately, and found inconsistency between the dipoles in the two surveys. These works argue that the analysis in \citet{2022ApJ...931L..14D}, which combines two catalogs, may be susceptible to systematic errors. Additionally, recent dipole measurements from quasar catalogs also find amplitudes exceeding CMB expectations at $\sim 5\sigma$ statistical significance \citep{2021ApJ...908L..51S,2023MNRAS.525..231D}. Indeed, a recent review article includes the discrepancy between the CMB and LSS dipole measurements in a list of possible anomalies in the standard cosmological model \citep{2022AnPhy.44769159P}. Besides exploring new physics to explain the observations, these discrepancies also invite further studies regarding
possible systematic errors in the measurements as well as additional efforts to model the dipole signals.

The LSS dipole measurements are indeed sensitive to various systematic and modeling uncertainties. On the measurement side, non-uniform sensitivity, large-scale calibration errors, survey masking, and the presence of foreground emission or extinction can introduce biases if not accounted for during the dipole estimation process. Therefore, dipole analyses must rigorously validate their data processing methodologies and ensure that the dipole signal estimators employed remain unbiased against all of these effects.

That said, there are also uncertainties in modeling the expected LSS dipole signals. First, as will be elaborated further in what follows, the kinematic LSS dipole depends on the
spectra of the source populations and their number counts as a function of flux, which may each evolve with redshift. This leads to uncertainties in predicting the expected kinematic LSS dipole from the CMB measurements. In the case of quasar studies, previous work has shown that these uncertainties may be important \citep{2022MNRAS.512.3895D,2022arXiv221204925G}, while their relevance for NVSS measurements has not yet been quantified.  

Second, and more important in the case of the NVSS catalog, the sources may be inhomogeneously distributed across the sky. That is, source clustering and Poisson fluctuations in the discrete NVSS source counts (i.e. shot noise) may each contribute to the measured dipole anisotropy. We refer to this as the ``local-source dipole,'' since distant sources are expected to be almost uniformly distributed, while the clustering of more nearby objects will give rise to a non-negligible dipole moment. We emphasize that the local-source clustering and shot-noise contributions have nothing to do with our motion relative to the LSS/CMB rest frame, and so constitute a nuisance for extracting the kinematic LSS dipole. A proper evaluation of the clustering term requires accurate knowledge of the redshift distribution of the LSS sources \citep{2012MNRAS.427.1994G,2016JCAP...03..062T,2018JCAP...04..031B,2023MNRAS.525..231D}.

The main objective of this work is to reevaluate the apparent discrepancy between the CMB and LSS dipoles, focusing specifically on NVSS radio dipole measurements. The NVSS catalog has been used extensively in previous radio dipole measurements, and a broad range of data processing schemes and dipole estimators have been employed in earlier studies. Consequently, here we mainly take current NVSS dipole estimates at face value. Instead of exploring measurement systematics, our focus is on constructing improved models of the NVSS dipole. Specifically, our goal is to predict the expected NVSS dipole in the standard $\Lambda$CDM cosmological model, including the kinematic, clustering, and shot-noise contributions to the dipole measurements and associated uncertainties. Our comprehensive approach will determine the probability distribution in $\Lambda$CDM for measuring a given total NVSS dipole amplitude and direction, conditioned on the CMB dipole measurement (under the assumption that it is entirely kinematic in origin). Here our analysis has overlap with recent work by \citet{2023MNRAS.525..231D} who performed related calculations to help interpret their quasar dipole measurement. Our work differs from this previous study since we consider NVSS dipole measurements, rather than the quasar case, while our approaches also differ somewhat. However, we have verified that our formulas agree with those in this earlier work where there is direct overlap. 

In order to model the clustering dipole, one needs to know the redshift distribution of the source populations and their biasing with respect to the underlying matter distribution. Many previous studies have relied on assumptions regarding the functional form of the NVSS redshift distribution and the redshift evolution of the source clustering bias \citep{2016JCAP...03..062T, 2018JCAP...04..031B}. Here, we improve on previous estimates of the clustering dipole by using a combination of the clustering redshift method \citep{2015MNRAS.446.2696S,2019ApJ...870..120C,2019ApJ...877..150C} and a cross-matching technique between NVSS and spectroscopic redshift surveys. The clustering redshift technique exploits cross-correlations between the NVSS data and external spectroscopic catalogs of galaxies and quasars with known redshifts. On large-scales where linear perturbation theory is applicable, this cross-correlation determines, statistically, a product between the NVSS redshift distribution and clustering bias. This can then, in turn, be extrapolated to predict the angular power spectrum at $\ell=1$ and the NVSS clustering dipole. For the smallest redshifts in the NVSS sample, however, we find that a cross-matching analysis is preferable and use this to inform our clustering determinations. Taken together, we obtain more reliable estimates of the NVSS redshift distribution and the resulting clustering dipole than in previous work. We then use this new determination as input to our probabilistic model and test consistency between NVSS and CMB dipole measurements in $\Lambda$CDM.

This paper is organized as follows. Sec.~\ref{S:formalism} presents the probabilistic expression for the total measured dipole. Sec.~\ref{S:data_processing} describes our procedure for processing the NVSS data, which is necessary for building our models for the kinematic and local-source dipoles as detailed in Sec.~\ref{S:modeling}. In Sec.~\ref{S:results}, we present a model for the probability distribution of the total NVSS dipole and compare it with previous measurements. We also quantify a few remaining sources of model uncertainty and discuss future prospects in Sec.~\ref{S:discussion}. Finally, we conclude in
Sec.~\ref{S:conclusion}. We provide detailed derivations of the relevant formulas in the Appendix. Throughout this work, we assume a flat $\Lambda$CDM cosmology consistent with the measurements from Planck \citep{2020AA...641A...6P}.

\section{Formalism}\label{S:formalism}
This section introduces an expression for the probability distribution function (PDF) of the dipole of an LSS tracer, denoted as $\mathbf{d}$, given our velocity with respect to the CMB frame, $\mathbf{v}_k$, as inferred from the CMB dipole measurement. Note that both $\mathbf{d}$ and $\mathbf{v}_k$ are 3D vectors. We adopt the assumption that our universe follows a $\Lambda$CDM cosmology, where the CMB and LSS share the same rest frame, while $\mathbf{v}_k$ is derived from the CMB dipole, which is assumed to have a purely kinematic origin. Hence the PDF we are after describes the prior probability of measuring a dipole, $\mathbf{d}$, from the LSS in $\Lambda$CDM, conditioned on our knowledge of the CMB kinematic dipole. That is, we seek a model for $P(\mathbf{d}|\mathbf{v}_k)$. For brevity, we generally refer to this object in what follows as $P(\mathbf{d})$. Here, we present a concise overview of the relevant formalism and provide detailed derivations in Appendix~\ref{A:formalism}.

Strictly speaking, our PDF derivation assumes that the LSS survey covers the entire sky, $f_{\rm sky}=1$. In the more realistic case of partial sky coverage, the PDF may deviate somewhat from the idealized form considered here while mode-mixing effects may also leak power from higher multipole moments into the dipole. In the NVSS case considered here, however, the sky coverage is large, $f_{\rm sky} \sim 0.6$. We therefore adopt the simple approximation that the PDF maintains the same overall shape as in the full-sky model but with an enhanced variance to account for the leakage. See below and Sec.~\ref{S:partial_sky} for further discussion.

Some previous studies have compared their dipole measurements solely with the expected kinematic dipole induced by our motion relative to the CMB frame. As alluded to in the Introduction, here we additionally include source clustering and shot noise as these will also contribute to LSS dipole measurements. We derive the PDF of the total dipole $\mathbf{d}$ taking into account all three contributions (kinematic, clustering, and shot noise).  

The measured dipole, $\mathbf{d}$, is the vector sum of the kinematic dipole, $\mathbf{d}_k$, and the local-source dipole, $\mathbf{d}_r$,
\begin{equation}
\mathbf{d} = \mathbf{d}_k+ \mathbf{d}_r.
\end{equation}
The clustering term and the shot-noise term can be grouped together in what we refer to as the ``local-source term,''
\begin{equation}
\mathbf{d}_r = \mathbf{d}_{\rm clus}+ \mathbf{d}_{\rm SN}.
\end{equation}
Without loss of generality, we define $\mathbf{\hat{z}}$ as the direction of the kinematic dipole. Thus, we can express the kinematic dipole vector as $\mathbf{d}_k = d_{k,x} \mathbf{\hat{x}} + d_{k,y} \mathbf{\hat{y}} + d_{k,z} \mathbf{\hat{z}} = d_k \mathbf{\hat{z}}$, where $d_k$ is the kinematic dipole amplitude. Here we assume that the CMB dipole has a pure kinematic origin, and therefore our motion is aligned with the direction of the CMB dipole, as determined by Planck to within an accuracy of $30''$ \citep{2020AA...641A...1P}. The measurement uncertainty for the radio-source dipole of interest in this study is on the order of tens of degrees. Consequently, we can consider the kinematic dipole direction to be precisely known from the CMB and disregard its measurement uncertainty in our model.

Although the speed of our motion relative to the CMB frame, $v_k$, has been determined by Planck to $0.025\%$ precision \citep{2020AA...641A...1P}, the amplitude of the kinematic LSS dipole also depends on factors such as the spectral index, luminosity function, and redshift distribution of the observed sources. The uncertainty associated with $d_k$ arises owing to our imprecise knowledge of these factors, as discussed in detail in Section~\ref{S:kinematic_dipole}. To model the PDF of the kinematic dipole amplitude, we assume a Gaussian distribution with a mean of $\overline{d}_k$ and a variance of $\sigma_k^2$:
\begin{equation}\label{E:P(d_k)_1D}
P(d_k) = \frac{1}{\sqrt{2\pi\sigma_k^2}}e^{-\frac{(d_k-\overline{d}_k)^2}{2\sigma_k^2}}.
\end{equation}

The presence of the local-source dipole introduces additional uncertainties in the total measured dipole, $\mathbf{d}$. The local-source dipole, $\mathbf{d}_r$, follows a zero-mean Gaussian random distribution (Eq.~\ref{E:P(d_r)_3D}) with variances in each dimension (i.e., in each of $\mathbf{\hat{x}}$, $\mathbf{\hat{y}}$, and $\mathbf{\hat{z}}$), given by
\begin{equation}\label{E:sigma_r}
\sigma^2_r \approx \frac{1}{f_{\rm sky}}\frac{3}{4\pi}\left(C_1^{\rm clus} + C_1^{\rm SN}\right),
\end{equation}
where $C_1$ represents the angular power spectrum of the $\ell=1$ modes, and the $3/4\pi$ prefactor comes from the normalization of the spherical harmonics  (i.e., $Y_{10}=\sqrt{3/4\pi}\,{\rm cos}\,\theta$). The first term describes the clustering power, while the second term accounts for shot noise. This formula is exact in the full-sky limit ($f_{\rm sky}=1$), while the $f_{\rm sky}$ factor is an approximation intended to account for the likely increase in variance under partial sky coverage (see also \citealt{2021JCAP...11..009N}). Among other things, this neglects any anisotropy in the sky coverage. The partial-sky effects are discussed further in Sec.~\ref{S:partial_sky}.

With the PDFs of $\mathbf{d}_k$ (Eq.~\ref{E:P(d_k)_3D}) and $\mathbf{d}_r$ (Eq.~\ref{E:P(d_r)_3D}), we can derive the PDF of the total measured dipole $\mathbf{d}$. We express the PDF, $P(\mathbf{d})$, in terms of its amplitude, $d$, and its angle, $\theta$, with respect to the velocity $\mathbf{v}_k$ direction (aligned with $\mathbf{\hat{z}}$ in our definition, which also matches the direction of the CMB dipole). These two quantities -- the dipole amplitude and direction -- are commonly reported in the literature, allowing for a meaningful comparison between our model PDF and previous dipole measurements. For a given $\sigma_r$, the joint PDF of $d$ and $\theta$ is:
\begin{equation}\label{E:p(dth)}
\begin{split}
P(d,\theta|\sigma_r)=&\frac{{\rm sin}\,\theta}{\sqrt{2\pi\sigma_r^2(1+\kappa^2)}}\frac{d^2}{\sigma_r^2}\\
&\cdot e^{-\frac{(1+(1-\mu^2)\kappa^2)d^2-2\overline{d}_k\mu d+ \overline{d}_k^2}{2\sigma_r^2(1+\kappa^2)}},
\end{split}
\end{equation}
where $\kappa\equiv\sigma_k/\sigma_r$, $\mu\equiv {\rm cos}\,\theta$. 
The PDF here is also known as the Kent distribution \citep{kent1982fisher}. In addition, we can obtain the marginalized PDFs $P(d|\sigma_r)$ and $P(\theta|\sigma_r)$ by integrating over $\theta$ and $d$, respectively (Eq.~\ref{E:P(d)} and ~\ref{E:P(th)}). Here the PDFs indicate a conditioning on $\sigma_r$ because, in practice, we have an imperfect knowledge of the redshift distribution and this translates into uncertainty in our knowledge of $C_1^{\rm clus}$.

We can account for the impact of uncertainty in the redshift distribution using a PDF for $C_1^{\rm clus}$, $P(C_1^{\rm clus})$, which will be determined in Sec.~\ref{S:clus_dipole}. We can then derive the PDF of $\sigma_r$, $P(\sigma_r)$, with Eq.~\ref{E:P(sigmar)}. Our final expression for the joint and marginalized PDFs of $d$ and $\theta$ are given after marginalizing over $P(\sigma_r)$ as:
\begin{equation}\label{E:P_marg}
\begin{split}
P(d,\theta) &= \int d\sigma_r P(d,\theta|\sigma_r)P(\sigma_r),\\
P(d) &= \int d\sigma_r P(d|\sigma_r)P(\sigma_r),\\
P(\theta) &= \int d\sigma_r P(\theta|\sigma_r)P(\sigma_r).
\end{split}
\end{equation}
These PDFs describe the expected prior distribution for the total dipole amplitude and direction (relative to the CMB dipole) in our model, which can be compared to dipole measurements in the current literature. 

One simplification adopted here is that we have neglected any correlation between the kinematic and clustering dipoles. In fact, the kinematic and clustering dipoles may actually be partly correlated. This is because our motion relative to the CMB frame, and the kinematic dipole, are sourced by surrounding density inhomogeneities. Likewise, the inhomogeneous NVSS source distribution also partly traces the same density variations, although with a different weighting in redshift and wavenumber. Our baseline results neglect this correlation for simplicity, although we model it in Appendix \ref{A:formalism} and give a detailed discussion there and in Sec.~\ref{S:kin_clus_corr}.

In principle, another potential complication is from any bulk motion of the tracers in the LSS sample relative to the CMB frame. In practice, after averaging over the NVSS source distribution, which probes a large cosmological volume, we find this to be negligibly small. Hence, we assume that the NVSS sources share the CMB frame (i.e. have negligibly small motion relative to this frame in $\Lambda$CDM). We justify this approximation quantitatively in Sec.~\ref{S:bulk_flow}.

Here we take the NVSS dipole estimates in the literature at face value, and do not attempt to account for any additional systematic error contributions beyond those in the reported measurement errors.  Note that the choices of source selection and masking vary across previous measurements, and any biases likely depend on the specific estimator employed. However, extensive tests have been conducted to verify that the NVSS dipole measurements are free from bias \citep[e.g., ][]{2021AA...653A...9S}. Moreover, previously reported NVSS dipole measurements exhibit fairly good consistency with each other (see Table~\ref{T:dipole}). Hence, we will compare our probabilistic dipole model with a compilation of results in the literature ignoring any
additional systematic errors. 

\section{Data Processing}\label{S:data_processing}
Although our focus in this work is on modeling the expected NVSS dipole signal, we do analyze the NVSS data itself in order to inform our calculations of the clustering, shot-noise, and kinematic dipoles. We compare these estimates with previous NVSS dipole measurements. In order to model the local-source dipole, we leverage the small-scale auto correlations in the NVSS data and their cross-correlations with spectroscopic reference catalogs. We use the cross-correlations to infer the NVSS redshift distribution and apply linear perturbation theory to predict the NVSS auto-power spectrum. This allows estimates of the clustering and shot-noise contribution to the dipole. We validate the linear-theory prediction with a direct measurement of the NVSS auto-power spectrum at intermediate scales. We then extrapolate this model to the dipole mode ($\ell=1$) to estimate the expected local-source dipole. While the dipole estimates themselves may be susceptible to mode-mixing effects from the survey mask and to uncertainties in the NVSS selection function, our smaller-scale measurements should be less sensitive to these concerns.

To ensure a meaningful comparison, we will process the NVSS data in a similar way to previous dipole studies. While we acknowledge that masking, filtering, and source selection procedures vary across different studies, our model primarily relies on small-scale clustering information from the NVSS data. This reduces susceptibility to many systematic concerns associated with the dipole measurements themselves. Our data handling procedures also align with several commonly adopted choices in the literature. Furthermore, we have validated the robustness of our model by finding consistent results under various masking and source selection criteria.

\subsection{NVSS}
The NVSS covers the entire sky area north of a declination of $-40^{\circ}$, corresponding to a sky covering fraction of $f_{\rm sky}=0.82$, at a central frequency of 1.4 GHz and with an angular resolution of $45^{''}$. The large sky coverage makes the NVSS dataset well-suited for measuring the cosmic radio dipole signal. The full NVSS source catalog contains $\sim 1.7$ million sources with specific fluxes of $S_\nu > 2.5$ mJy.

\subsection{Source Selection}\label{S:source_selection}
The completeness of the NVSS catalog is known to depend on declination. This is because the VLA D configuration was used to observe the region between declinations of $-10^\circ$ and $78^\circ$, while the VLA DnC configuration was used for the remaining area. Previous studies have suggested that this declination dependence can be mitigated by applying a flux threshold of $\sim 15$ mJy \citep[e.g., ][]{2002MNRAS.329L..37B, 2023arXiv230515335W}. The $15$ mJy flux threshold is hence commonly used in NVSS dipole measurements, and we thus adopt it in our analysis. This flux threshold leaves $\sim330,000$ NVSS sources in our analysis (after applying the Galactic and ecliptic plane masks described in Sec.~\ref{S:masking}).

In previous NVSS dipole studies, various different methods are employed to mitigate the dipole anisotropy induced by the clustering of nearby radio sources. One common approach (which we will also consider along with other techniques) is to cross-match the NVSS sources with external catalogs, as this provides redshift estimates for some of the sources. By removing NVSS objects that match nearby galaxies in the external catalogs, the systematic biases from local-source dipole signals can be reduced \citep[e.g., ][]{2002Natur.416..150B,2017MNRAS.471.1045C}.

In this analysis, we aim to build our model with a radio source population similar to that used in previous NVSS dipole studies, and we thus restrict our clustering dipole estimates to sources at $z> 0.01$ (corresponding to a distance of $44$ Mpc). Below this redshift, the peculiar velocities of galaxies become significant compared to the Hubble flow (e.g. a typical peculiar velocity of $300$ km s$^{-1}$ is a 10\% correction to the Hubble recession speed at $z=0.01$, while such velocities become increasingly important more nearby),  leading to large uncertainties in distance estimates based on photometric and spectroscopic redshifts. While previous NVSS dipole measurements employed different criteria to remove local sources, and it is possible that some sources with $z< 0.01$ were included in previous NVSS dipole measurements, our conservative approach of assuming their removal provides a lower bound on the clustering dipole. In order to consistently remove nearby sources in our auto and cross-correlation analyses, we  
exclude all NVSS sources that have matches in either the 2MASS Redshift Survey \citep[2MRS; ][]{2012ApJS..199...26H} or the local radio sources (LRS) registered on NED\footnote{\url{https://ned.ipac.caltech.edu/}} with $z<0.01$. We use $45{''}$, the FWHM of the NVSS beam, as the matching distance. This procedure removes approximately $2\%$ of NVSS sources with fluxes $S_\nu >15$ mJy.

We found that the removal of nearby sources matched to 2MRS or LRS sources has a negligible impact on our estimates of the clustering and shot-noise power spectra in higher-$\ell$ modes. However, it is important to note that nearby sources can still introduce significant dipole anisotropy if they remain in the catalogs used for dipole estimates. In Section~\ref{S:clus_dipole}, we further quantify the potential contribution to the clustering dipole from structures at $z<0.01$.

\subsection{Masking}\label{S:masking}
The choice of masking criteria adopted in previous NVSS dipole measurements has varied. Here, we employ a common strategy used in the literature. It is worth noting that these choices appear to have little impact on the NVSS dipole estimates because previous results using a $15$ mJy flux threshold are largely consistent with each other, in spite of the different masks employed. Therefore, it seems unlikely that this is a critical systematic concern.

We use the Healpix pixelization scheme \citep{2005ApJ...622..759G} with $N_{\rm side}=1024$, corresponding to an angular resolution of $3'.44$. All masking operations are performed in this pixelization scheme. We mask out regions within a Galactic latitude of $\pm 10^\circ$, as this is a common masking choice in previous work \citep{2011ApJ...742L..23S, 2021AA...653A...9S}.

In some NVSS dipole measurements, certain extended regions with high source density are identified and masked out \citep[e.g., ][]{2002Natur.416..150B,2023arXiv230515335W}. To assess the impact of these extended regions on our modeling, we visually inspect the NVSS source count map and identify a few such regions. However, we find that masking these extended regions does not significantly affect the NVSS angular power spectrum (Section~\ref{S:NVSS_Cl}) which is used to build our clustering (Section~\ref{S:clus_dipole}) and shot-noise models (Section~\ref{S:SN_dipole}). Therefore, we do not incorporate additional masks for these regions.

Bright and extended radio sources can be misidentified as multiple objects in the NVSS catalog due to the sidelobes of the instrumental beam. \citet{2002MNRAS.329L..37B} quantify the fraction of these multi-component sources by analyzing the small-scale angular correlation function in NVSS. We will incorporate their results to assess and correct for the impact of multi-component sources on our shot-noise model (Sec.~\ref{S:SN_dipole}). Hence, we adopt the same masking strategy for bright sources as in \citet{2002MNRAS.329L..37B}, which is to mask a $0.5^\circ$ region around all $S_\nu>1$ Jy NVSS sources. 

In some NVSS dipole measurements, a mask is applied around the Supergalactic plane \citep[e.g., ][]{2015APh....61....1T, 2017MNRAS.471.1045C}. However, multiple studies demonstrate that NVSS dipole estimates are insensitive to  masking along supergalactic latitude \citep{2011ApJ...742L..23S,2012MNRAS.427.1994G, 2017MNRAS.471.1045C}. Moreover, our NVSS cross-correlation and auto-correlation measurements are mostly confined to $\ell \gtrsim 10$ and likely less affected by large-scale systematics. Therefore, we do not apply any masking or filtering around the Supergalactic plane in our analysis.

In summary, our masking region includes the NVSS footprint and Galactic plane masks, as well as the bright source mask. This results in a total sky coverage of $f_{\rm sky}=0.63$. Again, we emphasize that our primary use of the NVSS data is to estimate cross-correlations with spectroscopic reference catalogs as this helps us determine the expected clustering dipole signal (Sec.~\ref{S:clus_dipole}). Since this uses only the smaller-scale information in the NVSS catalogs, we should be less impacted by systematic concerns that may affect the dipole measurements themselves.

\section{Modeling}\label{S:modeling}
\subsection{Kinematic Dipole Modeling}\label{S:kinematic_dipole}
\citet{1984MNRAS.206..377E} first arrived at a simple formula for the kinematic dipole in a radio survey with a uniform flux threshold, where the number counts around the specific flux threshold follow a power law with slope $x$:
\begin{equation}
\frac{dN}{dS_\nu}(>S_\nu)\propto S_\nu^{-x},
\end{equation}
and all radio sources have a power-law spectral index $\alpha$,
\begin{equation}
S_\nu \propto \nu^{-\alpha}.
\end{equation}
The kinematic LSS dipole signal arises from the combined effects of Doppler shifts and relativistic aberration. It can be expressed as:
\begin{equation}
\label{E:dkin}
d_k = A\beta = [2 + x(1+\alpha)]\beta,
\end{equation}
where $A=2 + x(1+\alpha)$, $\beta=v_k/c$, and $v_k$ represents the velocity of our own (heliocentric) frame relative to the CMB rest frame. This motion is well-determined from the CMB dipole measurements under the usual assumptions. This expression is widely
used in the subsequent literature on radio dipoles. As mentioned earlier, there are potential concerns with this estimate, as it assumes that the spectral index is uniform across the sky and it ignores any redshift evolution in $\alpha$ and $x$. This has been explored in the context of quasar dipole measurements \citep{2023MNRAS.525..231D, 2022MNRAS.512.3895D, 2022arXiv221204925G}.
In addition, the power-law source count assumption may also be imperfect. Indeed, it is well-known that the NVSS flux density function cannot be completely described by a simple power-law function \citep{2013AA...555A.117R, 2018JCAP...04..031B, 2015APh....61....1T}, although -- as described below -- corrections to the power-law case have a negligible impact on our main conclusions.

In order to best account for these issues, we aim to construct a reasonable PDF for $d_k$. Here, this PDF is intended to describe the prior probability for the NVSS kinematic dipole given the CMB measurements. To estimate this, we model both the ensemble-average values of $x$ and $\alpha$ and their covariance. The ensemble here should be thought of as a collection of many NVSS-like surveys, with statistical properties matching those of the actual NVSS sample. We can then estimate the expected average kinematic dipole and the variance across the ensemble and use this to inform our choice of prior. 

More specifically, we follow here the approach of \citet{2023MNRAS.525..231D}. First note that the mean value of $A$ is given by $\overline{A}=2+\overline{x}(1+\overline{\alpha})+\rho\sigma_x\sigma_\alpha$, where $\sigma_x^2$ and $\sigma_\alpha^2$ are the variance of $x$ and $\alpha$, respectively, and $\rho$ is their correlation coefficient. As mentioned above, $x$ and $\alpha$ may evolve with redshift while correlations between the fluctuations in $x$ and $\alpha$
may be important.  The $\rho \sigma_x \sigma_\alpha$ term here is intended to account for these correlations. For the spectral index, $\alpha$, we adopt a mean value of $\overline{\alpha}=0.75$ and an rms value of $\sigma_\alpha=0.25$, consistent with those in \citet{2021AA...653A...9S}. These values are also supported by \citet{2019RAA....19...96T}, who constrained the distribution of radio spectral indices using common sources observed in the TGSS \citep[150 MHz; ][]{2017A&A...598A..78I} and NVSS \citep[1.4 GHz; ][]{1998AJ....115.1693C} surveys. 
In our model, we set $\overline{x}=1.04$, which corresponds to the ``mask d'' case value used in \citet{2021AA...653A...9S} with the same flux threshold of 15 mJy.\footnote{Note that the NVSS flux density distribution is not perfectly described by a power law, as mentioned above. Using, however, the alternative form suggested in \citet{2021AA...653A...9S} -- with a flux-dependent value of $x$ -- leads to a $\sim 10\%$ reduction in the kinematic dipole estimate. We have explicitly checked that this leads to negligible changes in our final results.} In order to estimate $\sigma_x$ we follow \citet{2023MNRAS.525..231D} in adopting the least informative prior for the PDF of $x$ (i.e. this is chosen in accordance with the principle of maximum entropy; \citealt{jaynes03}). Given the constraint of the known mean, $\overline{x}$, and the requirement that $x \geq 0$, this yields $\sigma_x=\overline{x}$. Since, a priori, the only constraint on $\rho$ is that it lies within the range of $[-1,1]$, we adopt a conservative estimate and set $\overline{A}=2+\overline{x}(1+\overline{\alpha})=3.82$ and $\sigma_A = \sigma_x\sigma_\alpha=0.26$, which takes the maximum possible value of $|\rho|=1$; i.e., we adopt an uniformative prior on the correlation coefficient. Note that we assume a Gaussian distribution for $P(d_k)$ (Eq. ~\ref{E:P(d_k)_1D}). For the velocity, $v_k$, we use the latest measurements from Planck: $v_k = 369.82\pm0.11$ km s$^{-1}$ \citep{2020AA...641A...1P}. This corresponds to a mean value of $\overline{\beta}=1.23\times 10^{-3}$ and a standard deviation of $\sigma_\beta=3.70\times10^{-7}$. With this, we can calculate the mean and variance of the kinematic dipole: $\overline{d}_k = \overline{A}\overline{\beta} = 0.471\times10^{-2}$ and $\sigma_k = \overline{d}_k\sigma_{A}/A=0.032\times10^{-2}$ (the uncertainty on $\beta$ is negligible compared to that on $A$). These values are used in our model for the PDF of the kinematic dipole (Eq.~\ref{E:P(d_k)_1D}).

Note that the adopted $7\%$ model uncertainty on $\overline{d}_k$ has a negligible impact on our estimates of the total dipole PDF. Explicitly, the width of the kinematic dipole distribution enters Eq.~\ref{E:p(dth)} as an extra multiplicative term, $1+\kappa^2=1+\sigma^2_k/\sigma^2_r$. Here, $\sigma_r$ is set by the amplitude of the clustering plus shot-noise contributions. As we will see in the following section, we find that the clustering plus shot-noise terms are comparable in amplitude to the expected kinematic dipole (i.e., to $\overline{d}_k$), and so $\kappa^2$ amounts to a negligible sub-percent correction.

\subsection{Clustering Dipole Modeling}\label{S:clus_dipole}
The amplitude of the clustering dipole, $d_{\rm clus}$, is related to  the angular power spectrum of the NVSS source fluctuations (at $\ell=1$) by $d_{\rm clus} = \sqrt{9 \, C_1^{\rm clus}/4\pi}$.
The clustering power spectrum is in turn modeled as:
\begin{equation}\label{E:Clclus}
C_\ell^{\rm clus} = \int \frac{dk}{k}\frac{2}{\pi}k^3 P(k) \left[\int dz \, b(z)\frac{1}{N}\frac{dN}{dz}D(z)j_\ell[k\chi(z)]\right]^2,
\end{equation}
where $P(k)$ is the matter power spectrum at $z=0$ according to linear theory, $b(z)$ is the clustering bias of the sources, $\frac{dN}{dz}$ is the surface density of sources per unit redshift, $N=\int dz \frac{dN}{dz}$ is the projected source surface density, $D(z)$ is the linear growth factor, $\chi$ is the comoving distance, and $j_\ell$ denotes a spherical Bessel function. Since we are mostly interested in the $\ell=1$ dipole mode, which is determined by low-$k$ fluctuations, adopting the linear theory power spectrum and assuming scale-independent biasing should be excellent approximations.

In order to calculate $C_\ell^{\rm clus}$, we require the redshift distribution, $(1/N) \, dN/dz$, and the clustering bias, $b(z)$, each as a function of redshift. Previous NVSS dipole studies have employed various approaches for modeling these quantities and their redshift evolution. For instance, \citet{2016JCAP...03..062T} inferred $\frac{dN}{dz}$ at the NVSS frequency using the source distribution in two other radio surveys (CENSORS and Hercules) and fit a parameterized $b(z)$ function to the NVSS angular auto-power spectra. \citet{2018JCAP...04..031B} modeled $\frac{dN}{dz}$ from simulations and adopted the same model for $b(z)$ as in \citet{2016JCAP...03..062T}.

However, a more direct and reliable approach is to use the clustering redshift technique \citep{2008ApJ...684...88N,2015MNRAS.446.2696S,2019ApJ...870..120C,2019ApJ...877..150C}. This method allows one to determine $b(z) \frac{dN}{dz}$ tomographically by cross-correlating the NVSS source catalog with external galaxy or quasar survey data in various spectroscopic redshift bins. In fact, \citet{2008PhRvD..78d3519H} previously applied this methodology to the NVSS data in a different context, while studying the integrated Sachs--Wolfe effect in the CMB. However, the external reference catalogs have improved a great deal since this application (e.g., their reference samples only covered up to $z=2$ and with limited redshfit resolution). We hence update these early calculations here. 

\begin{figure}[ht!]
\begin{center}
\includegraphics[width=\linewidth]{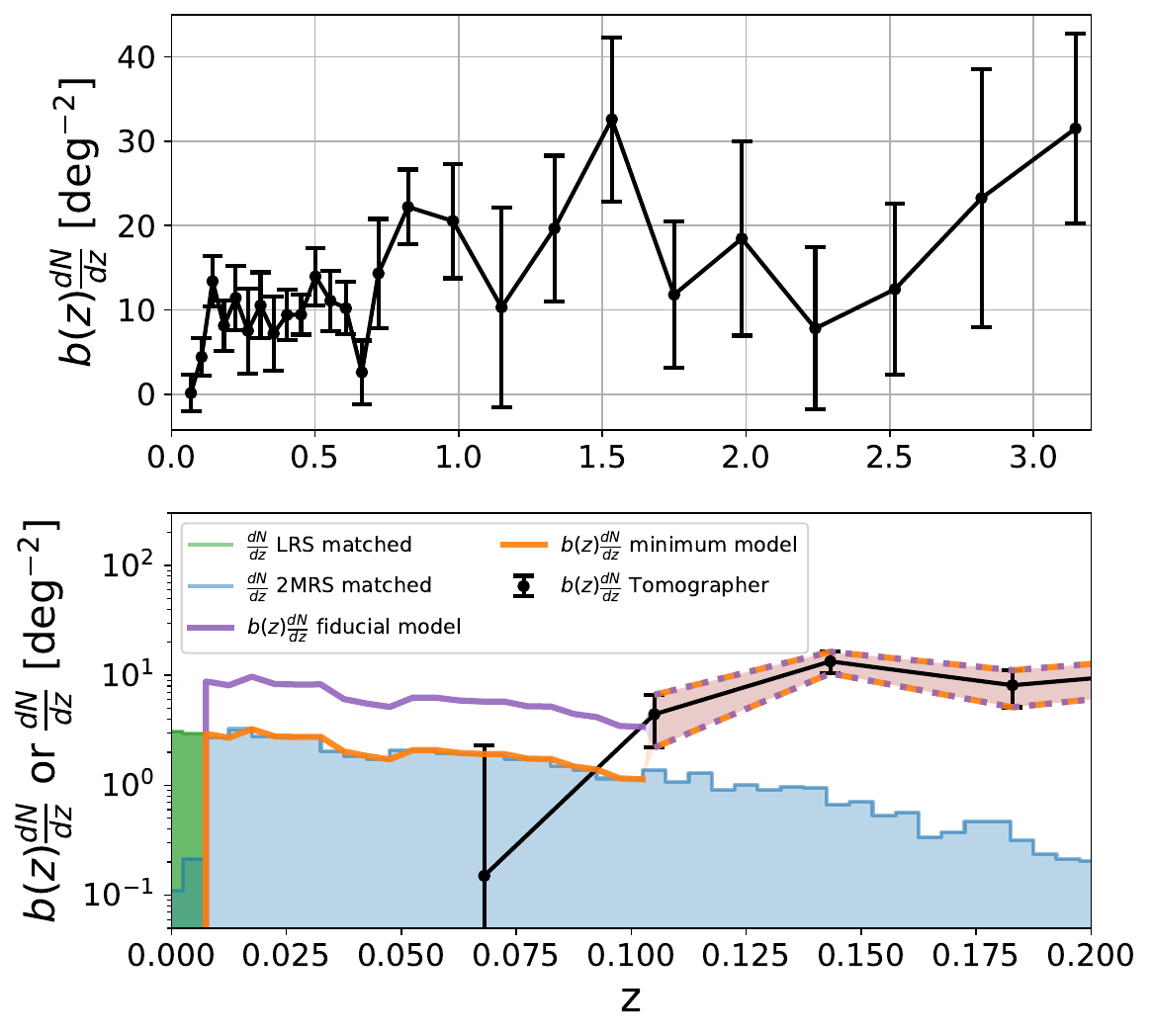}
\caption{\label{F:dNdz} Top: The redshift kernel, $b(z)\frac{dN}{dz}$, of NVSS sources inferred from \textit{Tomographer}. Bottom: The redshift distribution, $\frac{dN}{dz}$, of NVSS sources cross-matched to the 2MRS (blue) and LRS (green) catalogs. This is used to estimate the redshift distribution more nearby at $0.01 < z < 0.1$. The black data points show $b(z)\frac{dN}{dz}$ from \textit{Tomographer}, and are identical to those in the top panel. The orange and purple lines illustrate our minimum and fiducial models for $b(z)\frac{dN}{dz}$, respectively. Both models are cut-off at $z=0.01$ as we only model the clustering dipole at $z>0.01$. At $0.01 < z < 0.1$, the minimum model takes $\frac{dN}{dz}$ from the cross-match between NVSS and 2MRS as  $b(z)\frac{dN}{dz}$, while the fiducial model scales the minimum model upwards by a factor of 3 to match the \textit{Tomographer} data point at $z\sim 0.1$ (see text for a discussion). For $z> 0.1$ (the second \textit{Tomographer} data point), both models transition to using the \textit{Tomographer} results. To construct the PDF of $C_\ell^{\rm clus}$, we sample Gaussian realizations from the mean and variance at each redshift bin and linearly interpolate between them. The orange band shows the $1\sigma$ range where this sampling is performed.}
\end{center}
\end{figure}

Specifically, we use the publicly available \textit{Tomographer} package\footnote{\url{http://tomographer.org/}} to perform the clustering redshift calculations. This code infers $b(z) \frac{dN}{dz}$ (or related quantities in the case of a diffuse map as opposed to a discrete source catalog) for any given 2D projected map by cross-correlating it with pre-compiled galaxy and quasar samples from the Sloan Digital Sky Survey \citep{2005AJ....129.2562B,2016MNRAS.455.1553R,2018AA...613A..51P}. The resulting NVSS $b(z)\frac{dN}{dz}$ obtained from \textit{Tomographer} is displayed in the top panel of Fig.~\ref{F:dNdz}. This figure illustrates the rather broad distribution of NVSS source redshifts. While many of the NVSS objects are extremely distant, with the inferred distribution including sources all the way out to $z \gtrsim 3$,\footnote{Note that the \textit{Tomographer} reference samples only cover up to $z = 3.15$, but sources at such high redshifts contribute negligibly to the clustering dipole and so this incompleteness is unimportant for present purposes.} there are also a sizeable number of more nearby objects. This reflects the range of source populations included in the NVSS catalog, which contains both powerful active galactic nuclei (AGN) in the distant universe and also more nearby, yet less luminous star-forming galaxies (with radio emission powered mainly by synchrotron radiation from relativistic electrons accelerated in supernova explosions) \citep{2016era..book.....C}. While the distant objects in the NVSS sample span a large cosmological volume and are expected to be nearly uniformly distributed across the sky, the clustering of the nearby sources in the survey may nonetheless contribute to the dipole measurements. 

However, the lowest redshift bin in \textit{Tomographer} only reaches down to $z=0.06$ and the measurement in this particular bin is a non-detection (with NVSS). Given the sensitivity of the clustering dipole to nearby sources, we need to extend our redshift distribution model to lower redshifts. To do this, we perform a cross-match between the NVSS catalog and 2MRS, using a matching radius of $45{''}$, with the matched sources informing our model for the low-redshift end of the redshift distribution. The resulting $\frac{dN}{dz}$ distribution (for the matched sources) is shown in the bottom panel of Fig.~\ref{F:dNdz}. For further reference, we also show the number density of sources matched to 2MRS and LRS at $z < 0.01$, although these sources have been excluded in our data processing (Sec.\ref{S:source_selection}). Note that 2MRS does not provide a complete record of sources with $z< 0.01$, hence we utilize LRS as a complementary dataset. We note that this cross-matching procedure does not guarantee capturing all $z<0.1$ sources in the NVSS. This is because, first, there may be low-redshift radio sources in NVSS that do not have near-infrared counterparts in 2MRS. Second, NVSS is known to misidentify the center of some extended radio sources. Some previous works addressed the latter issue by cross-matching with higher-resolution radio surveys \citep{2005MNRAS.362....9B,2010ApJ...723.1119L,2018AJ....155..188L}. However, for present purposes, the cross-matching results are used only in setting a reasonable lower bound on the low-redshift NVSS source redshift distribution, and we thus leave more involved analyses to possible future work.

Based on the cross-matching results, we consider two models for $b(z) \frac{dN}{dz}$ at $0.01 < z < 0.1$. First, the ``minimum model'' uses the $\frac{dN}{dz}$ directly from the cross-match, which represents a lower limit on the clustering signal. This provides a conservative lower bound on $b(z) \frac{dN}{dz}$ both because there may be low-redshift radio sources in NVSS that do not have near-infrared counterparts in 2MRS, or where their extended nature in NVSS makes them hard to cross-match in 2MRS, and second because the clustering bias $b(z)$ is expected to exceed unity. In our ``fiducial model'' we instead multiply the redshift distribution in the minimum model by a factor of 3. This fiducial case reproduces the value of $b(z)\frac{dN}{dz}$ from \textit{Tomographer} at $z=0.1$ and is intended to account for the clustering bias neglected in the minimum model, and any incompleteness in the cross-matched sample. 

In summary, we use both \textit{Tomographer}/cross-matched models to predict the clustering dipole, and account for  uncertainties in the redshift distributions through the $C_\ell^{\rm clus}$ PDF (see the text around Eq.~\ref{E:P_marg}). Specifically, for each \textit{Tomographer} data point, we draw from a Gaussian realization with the specified mean and variance from \textit{Tomographer}, and perform linear interpolation between redshift bins. The error bars in \textit{Tomographer} are derived from bootstrapping, and our sampling procedure ignores the correlations between redshift bins, which should be a good approximation given the large separation in radial distance between most of the redshift bins.

We can also compare our clustering redshift/cross-matching results to previous parametric models for $b(z) \frac{dN}{dz}$ from the literature. Specifically, an explicit comparison with the earlier study of \citet[][hereafter NT15]{2015ApJ...812...85N} is given in Appendix~\ref{A:NT15_comparison}. There, we show that $b(z) \frac{dN}{dz}$ from NT15 generally lies within the range between our minimal and fiducial models, despite the different approach adopted in that work for calibrating the NVSS redshift distribution and clustering bias. The resulting total dipole PDF is consistent with the range bracketed by our minimal/fiducial estimates (see Appendix~\ref{A:NT15_comparison}, Figures~\ref{F:dNdz_NT15} and \ref{F:Pd_NT15} for details). However, as discussed further in Appendix~\ref{A:NT15_comparison}, \citet{2016JCAP...03..062T} use the NT15 clustering model to assess consistency with their own NVSS dipole measurements and find a larger difference with model expectations than found here. For the same flux cut (15 mJy), those authors (using a different approach) find a $\sim 2\sigma$ difference with their model, while we find $\sim 0.5\sigma$ agreement between their measurement and our own model. As further discussed in Appendix~\ref{A:NT15_comparison}, it is unclear why NT15 found a larger difference despite their similar estimates for the clustering and shot-noise dipoles. We provide a cross-check on our significance estimate in the Appendix~\ref{A:NT15_comparison}.

\begin{figure}[ht!]
\begin{center}
\includegraphics[width=\linewidth]{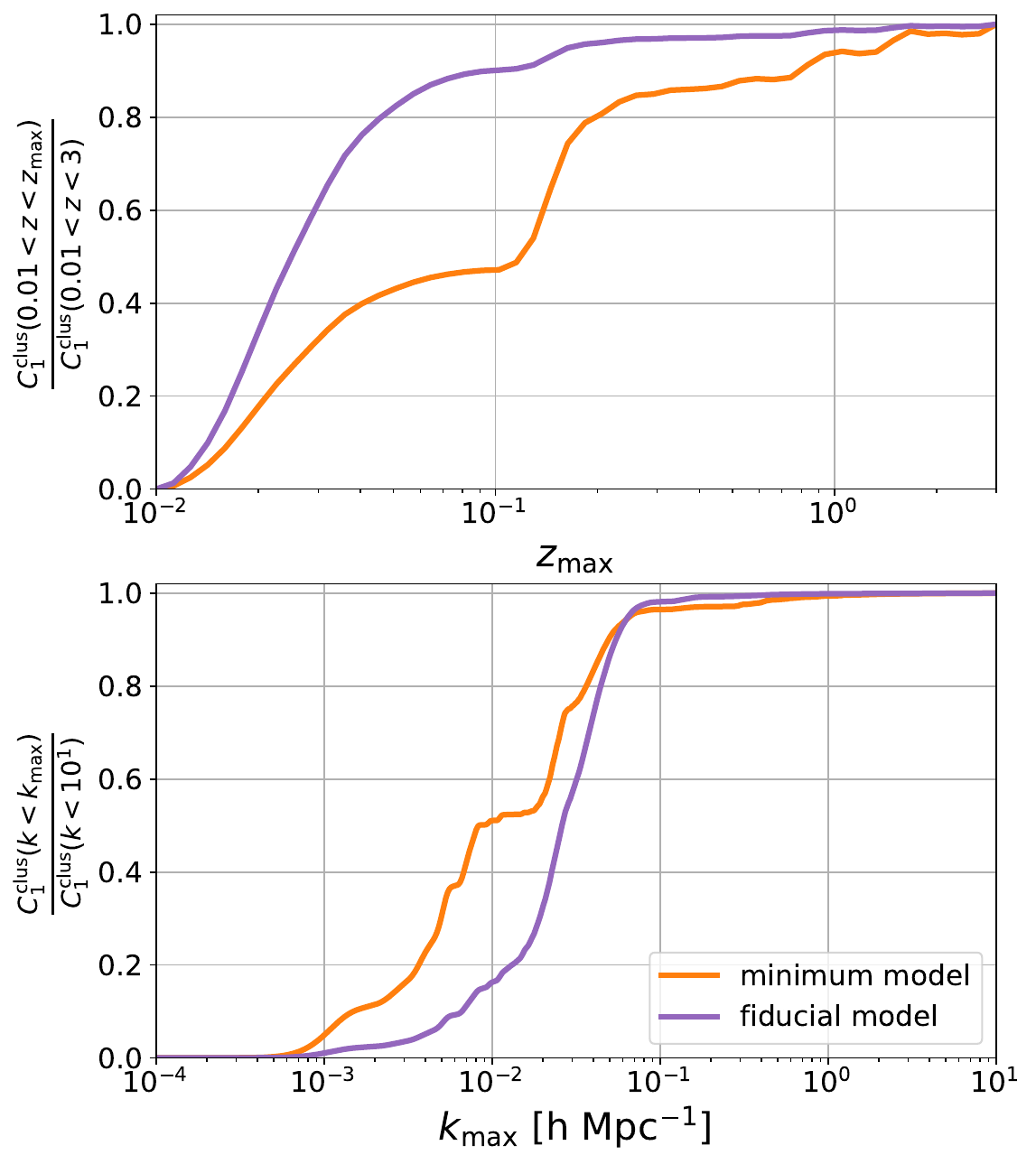}
\caption{\label{F:kz_kernel} The cumulative $C_1^{\rm clus}$ for the minimum (orange) and fiducial (purple) models as a function of the maximum redshift $z$ (top) and the maximum $k$ mode (bottom).}
\end{center}
\end{figure}

It is also worth noting that although the error bars on $b(z) \frac{dN}{dz}$ are still sizable, these uncertainties do not drive the width of the dipole amplitude PDF. Instead, this reflects the range of possible alignment angles between the kinematic, clustering, and shot-noise dipoles, and the cosmic/sample-variance in the clustering/shot-noise terms. Also, since the clustering dipole is dominated by local sources at $z\lesssim0.2$ (Fig.~\ref{F:kz_kernel}), the larger statistical uncertainties in $b(z) \frac{dN}{dz}$ at high redshift have little impact on our dipole estimates. We have explicitly checked that the statistical uncertainties on $b(z) \frac{dN}{dz}$ have negligible impact on our dipole PDF model.

The upper limit for the redshift integration in Eq.~\ref{E:Clclus} is set to $z=3$ as we have confirmed that the clustering dipole has negligible contributions from $z>3$ (see Fig.~\ref{F:kz_kernel}). For $z<0.1$, we assume the redshift distribution from the fiducial or minimum model built from cross-matching, and set the lower limit in the redshift integration of Eq.~\ref{E:Clclus} to $z=10^{-3}$. We assume that all of the local NVSS sources at still lower redshifts have been removed (see Sec.~\ref{S:source_selection}). With the PDF, $P(C_\ell^{\rm clus})$, determined through numerically sampling the \textit{Tomographer} results as described above, we can derive the PDF of $\sigma_r$ using Eq.~\ref{E:P(sigmar)}, which is used in Eq.~\ref{E:P_marg} to predict the total measured dipole.

Our minimum and fiducial models for the $\ell=1$ angular power spectrum yield rms clustering dipole amplitudes of $d_{\rm clus} =(0.20\pm0.07)\times 10^{-2}$ and $d_{\rm clus}=(0.43\pm0.07)\times 10^{-2}$, respectively. \citet{2018JCAP...04..031B} previously estimated the expected NVSS clustering dipole amplitude to be $d_{\rm clus}=0.23\times 10^{-2}$ using simulations. Their value falls between our minimum and fiducial estimates, despite the slight difference in the flux threshold ($20$ mJy) used in their analysis (we use $15$ mJy).

The top panel of Fig.~\ref{F:kz_kernel} shows the cumulative clustering dipole $C_1^{\rm clus}$ for both our minimum and fiducial models as a function of the maximum redshift, $z_{\rm max}$. For the minimum/fiducial model, approximately $80\%$ of the clustering dipole arises from structures below a redshift of $0.2/0.05$. The sharp transition in the minimum model at $z=0.1$ is a result of the discontinuity in our $b(z)\frac{dN}{dz}$ model (see Fig.~\ref{F:dNdz}). In a previous study by \citet{2016JCAP...03..062T}, the authors found that about $60\%$ of the clustering dipole originates from $z<0.1$ based on their model. This fraction falls between our minimum and fiducial models.

To assess the contribution of any $z< 0.01$ sources to the clustering dipole, which we neglect in our analysis, we extrapolate our two models to $z=10^{-3}$ using the cross-matching results, and calculate the clustering power spectrum, $C_\ell$, at $\ell=1$. Including structures from $10^{-3}<z<0.01$ increases the clustering power by approximately 40\% for both models, corresponding to a 20\% boost in the clustering dipole amplitude. While this contribution is not negligible, we will discuss in Sec.~\ref{S:discussion_srcselection_masking} how neglecting the nearby source contribution makes our clustering dipole prediction a lower estimate and does not significantly impact our main conclusions. 

For the dipole contributions from even more nearby sources at $z\lesssim 10^{-3}$, we can reasonably assume that most such local sources have been removed by our bright source mask at a threshold of $1$ Jy (Sec.~\ref{S:masking}). First, note that any such extremely nearby sources in the NVSS catalog will be star-forming galaxies rather than AGN.
To put the bright source threshold in context, then, note that there is a scaling relation between radio luminosity and star formation rate (SFR) in star-forming galaxies \citep{2018MNRAS.475.3010G}:
\begin{equation}
L_\nu=10^{29}\left[\frac{\rm SFR}{1\,M_\odot\, {\rm yr}^{-1}}\right]\left[\frac{\nu}{150\, {\rm MHz}}\right]^{-0.7} {\rm erg}\,{\rm s}^{-1}\,{\rm Hz}^{-1}.
\end{equation}
Here, we extrapolate to the NVSS frequency ($1.4$ GHz), and note that the $S_\nu >1$ Jy flux threshold corresponds to a galaxy at $z\sim 10^{-3}$ with a relatively small SFR of $\sim 1\,M_\odot$ yr$^{-1}$. That is, many extremely nearby sources will, in any case, be removed by the bright flux threshold. Note that while the $S_\nu >1$ Jy flux threshold has been used in some previous NVSS dipole measurements \citep[e.g., ][]{2016JCAP...03..062T}, other studies have employed different thresholds or alternative procedures for removing bright sources. Despite these variations, the overall agreement among reported dipole measurements in the literature suggests (see Table~\ref{T:dipole}) that the specific flux threshold used is not a critical factor in estimating the NVSS dipole given their measurement uncertainties. 

The bottom panel of Fig.~\ref{F:kz_kernel} displays the cumulative $k$-space window function for $C_1^{\rm clus}$. That is, it illustrates which wavenumbers produce most of the clustering dipole. In both the minimum and fiducial models, the clustering dipole power is predominantly contributed by modes in the range of $10^{-3}<k<10^{-1}$ h Mpc$^{-1}$. This is consistent with the findings of \citet{2016JCAP...03..062T}, who also studied the $k$-space window function of $C_1^{\rm clus}$ in their NVSS clustering model. Within this range of $k$, the matter power spectrum can be  well-described by linear theory. That is, we can neglect the effects of small-scale nonlinear clustering, which mainly enhance the matter power spectrum at $k\gtrsim 0.2$ h Mpc$^{-1}$ \citep{1994MNRAS.267.1020P,1996MNRAS.280L..19P}.  We also neglect large-scale relativistic correction terms, which primarily impact modes with $k<10^{-3}$ h Mpc$^{-1}$ \citep{2009PhRvD..80h3514Y}.

\subsection{Shot-noise Dipole Modeling}\label{S:SN_dipole}
The shot-noise contribution to the local-source dipole is also related to the shot noise in the angular power spectrum at $\ell=1$ via $d_{\rm SN}=\sqrt{9 \, C_1^{\rm SN}/4\pi}$. In general, the shot-noise angular power spectrum $C_1^{\rm SN}$ is given by the reciprocal of the source number density, $C_\ell^{\rm SN}=(dN/d\Omega)^{-1}$. That is, it is given by the inverse of the number of sources per steradian on the sky. However, some radio sources exhibit extended structures that exceed the NVSS beam size, resulting in multiple entries in the NVSS catalog \citep{2002MNRAS.329L..37B}. The presence of these multi-component sources changes the shot-noise level \citep{2004MNRAS.351..923B} according to:
\begin{equation}
\Delta C_\ell^{\rm SN} = \left(\frac{\overline{c^2}}{\overline{c}}-1\right)\left(\frac{dN}{d\Omega}\right)^{-1},
\end{equation}
where $\frac{dN}{d\Omega}$ is the number density of the sources in the NVSS catalog, which includes duplicate entries from multi-component sources. The values $\overline{c}=\langle c_i\rangle$ and $\overline{c^2}=\langle c^2_i\rangle$ represent the first and second moments of the number of components, $c_i$, per galaxy. In most cases, multi-component sources in NVSS are resolved into double sources. Assuming the fraction of these double sources is small, denoted as $e$ (where $e\ll 1$), we can approximate $\overline{c}=1+e$ and $\overline{c^2}=1+3e$ \citep{2004MNRAS.351..923B}. The value of $e$ has been measured as a function of the flux threshold in NVSS from small-scale two-point correlation functions by \citet{2002MNRAS.329L..37B}. For our flux threshold of 15 mJy, $e\sim0.08$, which leads to an effective shot noise of:
\begin{equation}
C_\ell^{\rm SN} = \frac{\overline{c^2}}{\overline{c}}\left(\frac{dN}{d\Omega}\right)^{-1} = 1.15 \cdot \left(\frac{dN}{d\Omega}\right)^{-1}=2.9\times10^{-5},
\end{equation}
corresponding to a shot-noise dipole of $d_{\rm SN}=0.46\times 10^{-2}$. 

In practice, the multi-component correction factor is only appropriate on scales larger than the angular sizes of the sources. \citet{2002MNRAS.329L..37B} found that, on average, this affects the correlations at $\theta\lesssim 0.5$ deg. However, some sources may still be partly resolved by $\ell\sim100$, which means at this scale, the shot-noise correction factor should be smaller than our model ($C_\ell^{\rm SN} = 1.15 (dN/d\Omega)^{-1}$). In fact, if we let the amplitude of the shot-noise vary and determine the best-fit correction factor to the NVSS auto-power spectrum measurements (black data points in Fig.~\ref{F:NVSS_Cl}), we find  $C_\ell^{\rm SN}=1.06(dN/d\Omega)^{-1}$. However, using this alternative value in the following analysis has a negligible impact on our final conclusions, and hence remaining shot-noise uncertainties seem unimportant. Since we are mainly concerned with the power at $\ell=1$, we use the unresolved value of $C_\ell^{\rm SN} = 1.15 (dN/d\Omega)^{-1}$ in what follows, but our results are insensitive to this particular choice. 

\subsection{Model Validation}\label{S:NVSS_Cl}
\begin{figure}[ht!]
\begin{center}
\includegraphics[width=\linewidth]{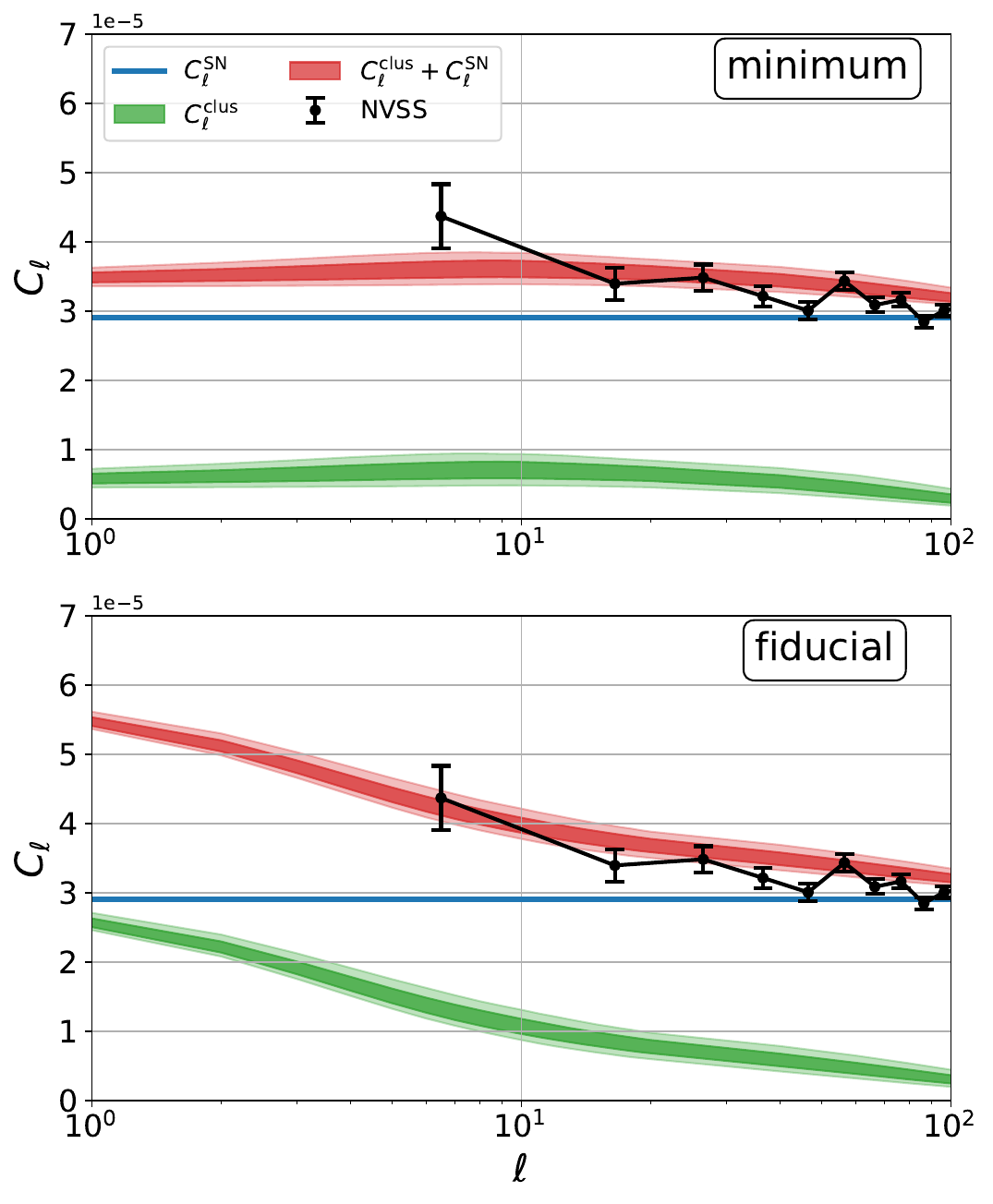}
\caption{\label{F:NVSS_Cl} A comparison between the NVSS auto-power spectrum and our clustering plus shot-noise models at intermediate $\ell$.
Black: angular power spectrum of the NVSS source count map. Blue: the shot-noise power spectrum. Green: the clustering power spectrum. Red: the local-source power spectrum given by summing the clustering and shot-noise terms. The dark and light shaded regions are the $68\%$ and $95\%$ confidence intervals, respectively, after accounting for redshift distribution uncertainties. The top and the bottom panels display the minimum and the fiducial clustering power spectrum models, respectively. Both panels have the same shot-noise (blue) and NVSS data (black) power spectra.}
\end{center}
\end{figure}

We validate our clustering and shot-noise models by comparing them to the NVSS auto-power spectrum at higher $\ell$. We calculate the angular power spectrum of the NVSS source count map after applying our source selection and masking criteria, as described in Section~\ref{S:data_processing}. We use the Healpix pixelization scheme with a resolution of $N_{\rm side}=1024$ to compute the power spectrum. To account for the effects of masking, we use the pseudo-$C_\ell$ estimator implemented in the NaMaster package \citep{2019MNRAS.484.4127A}. Since we are using the angular auto-power spectrum of NVSS sources only to cross-check our linear theory model we show results with $\ell > 6$ where cut-sky effects are not severe. The results are shown in Fig.~\ref{F:NVSS_Cl}. The consistency between our clustering plus shot-noise power spectrum models and the observed NVSS power spectrum in these higher $\ell$ modes helps validate our model for the NVSS redshift distribution within the assumed $\Lambda$CDM cosmological model. Furthermore, this also indicates that the NVSS source clustering estimates are not too contaminated by systematics, at least on the scales probed here. This consistency gives us confidence that we can robustly extrapolate our model to predict the local-source dipole at $\ell=1$.

Fig.~\ref{F:NVSS_Cl} illustrates that the clustering power at $\ell=1$ is comparable to the shot-noise in our fiducial case, while the clustering is approximately five times smaller in our minimum model. Note that $b(z)dN/dz$ at $z<0.1$ is 3 times smaller in the minimum model than in the fiducial case, and hence the clustering contribution from $z<0.1$ sources is $\sim 9$ times smaller, while the clustering dipole from higher redshift ($z>0.1$) sources is the same in both models. At $\ell=1$, the total source power, which is the sum of clustering and shot noise, is $C_1 = (3.5\pm0.07)\times 10^{-5}$ for the minimum model and $C_1 = (5.5\pm0.07)\times 10^{-5}$ for the fiducial model. This corresponds to a local-source dipole of $d_r=(0.50\pm0.07)\times10^{-2}$ for the minimum model and $d_r=(0.63\pm0.07)\times10^{-2}$ for the fiducial model. Note that the local-source dipole is comparable to the expected kinematic dipole, $\overline{d}_k = 0.471\times10^{-2}$, emphasizing the necessity of including contributions from shot-noise, clustering, and the kinematically induced dipole when comparing to NVSS measurements. Note that the similarity between the different contributions here is entirely coincidental as the terms have disparate origins, yet they nevertheless happen to roughly match for the NVSS sample.  

\begin{figure}[ht!]
\begin{center}
\includegraphics[width=\linewidth]{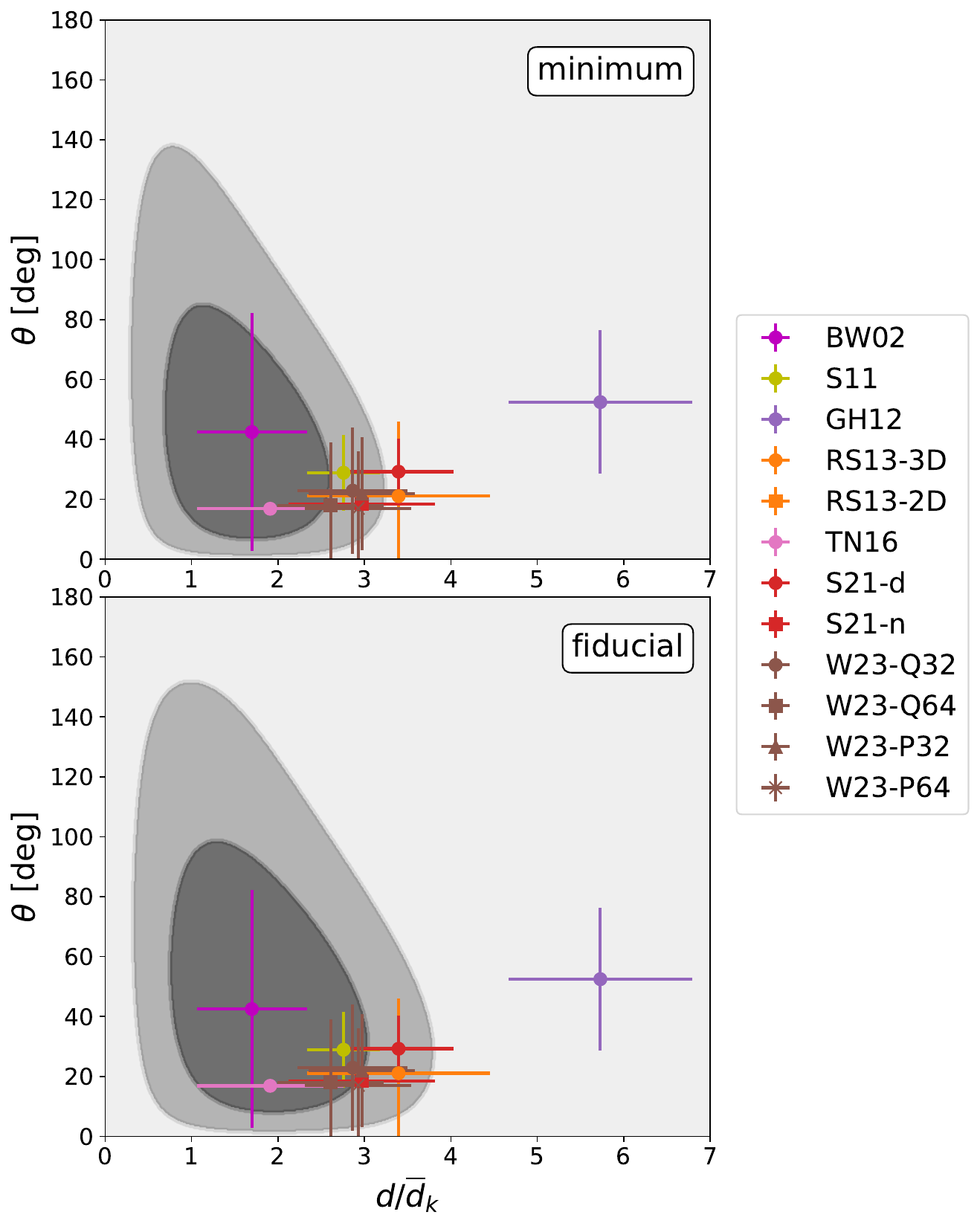}
\caption{\label{F:Pdth} 
The 2D PDF, $P(d,\theta)$, for the expected dipole amplitude and direction (relative to the CMB dipole) compared to NVSS measurements in the literature for the fiducial (top) and minimum (bottom) models. The model PDFs account for kinematic, clustering, and shot-noise contributions to the dipole measurements.  The dark and light gray areas indicate the expected dipole amplitude and direction at $68\%$ and $95\%$ confidence. The colored data points show NVSS dipole measurements obtained with a 15 mJy flux threshold (see Table~\ref{T:dipole} for more details). The dipole amplitudes are given in units of the expected (mean) kinematic dipole, i.e. we consider $d/\overline{d}_k$. For some data points in this figure and the following ones, the reported measurement errors partly arise from the shot-noise and/or clustering contributions that also set the width of the model PDFs and care must be taken to avoid ``double-counting'' portions of the error budget (see text for details).}
\end{center}
\end{figure}

\section{Results}\label{S:results}
With our models for the kinematic and local-source dipoles presented in Sec.~\ref{S:modeling}, we compute the PDF of the expected NVSS dipole amplitude and angle (relative to the CMB dipole), $P(d,\theta)$, as given by Eq.~\ref{E:p(dth)} and \ref{E:P_marg}. The results of our minimum and fiducial models are shown in Fig.~\ref{F:Pdth}. We compare our PDF with various NVSS dipole measurements obtained at the same flux threshold of $15$ mJy, as are summarized in Table~\ref{T:dipole}\footnote{Among the measurements listed in Table~\ref{T:dipole}, only \citet{2021AA...653A...9S} directly provide the offset angle, $\theta$, relative to the CMB dipole, along with its uncertainty, $\sigma_\theta$. The remaining measurements report the coordinates of their dipole directions in R.A. ($\alpha$) and Dec. ($\delta$). Hence, we calculate $\theta$ as the angular separation between the mean value of these coordinates and the CMB dipole direction given by \citep{2020AA...641A...1P}, and their $\sigma_\theta^2$ uncertainties using $\sigma_\theta^2 = \sigma_{\delta}^2+\sigma_{\alpha}^2{\rm cos}^2{\delta}$.}.

\begin{figure}[ht!]
\begin{center}
\includegraphics[width=\linewidth]{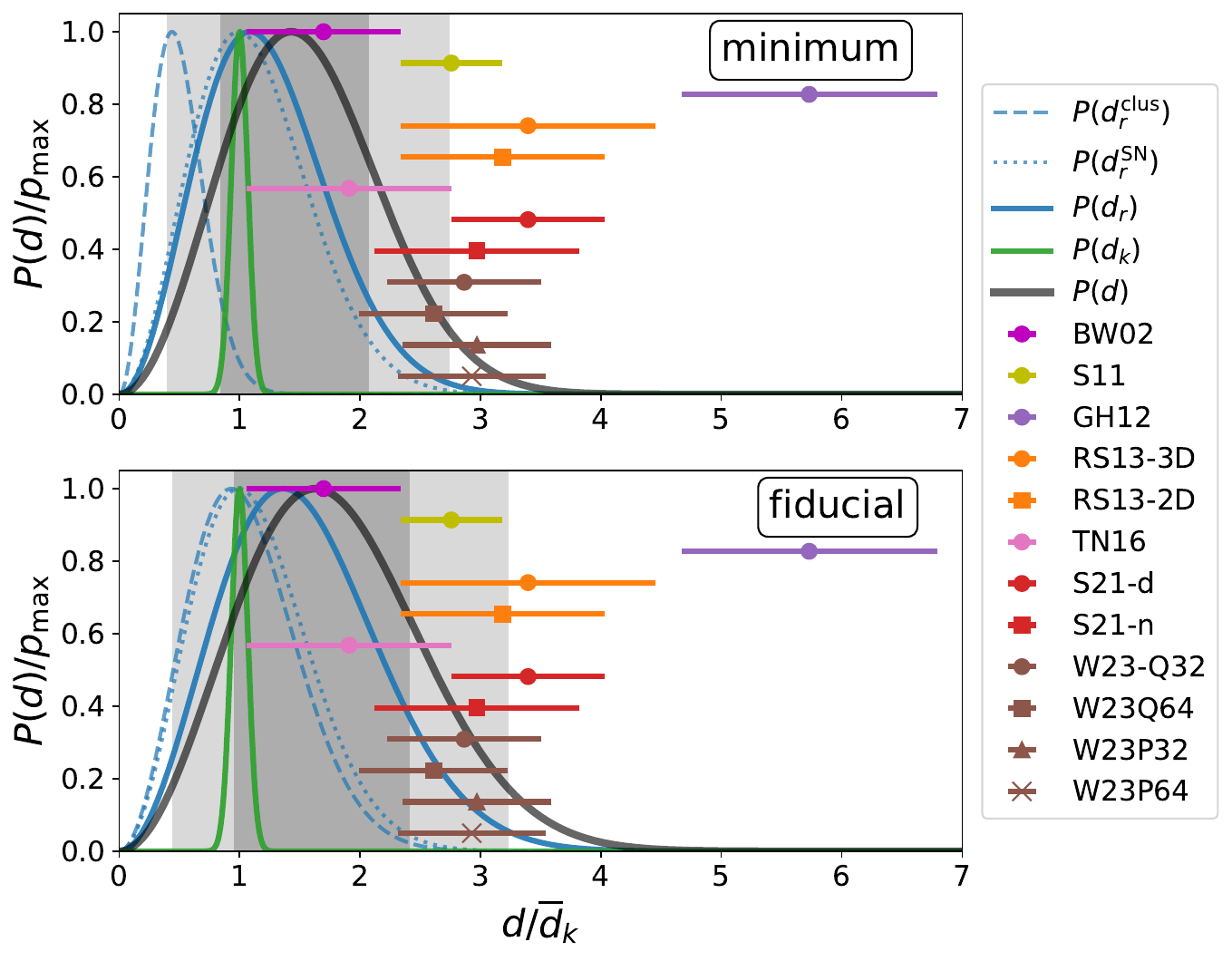}
\caption{\label{F:Pd} The 1D marginalized PDF for the expected dipole amplitude $P(d)$ (black; Eq.~\ref{E:P(d)} and Eq.~\ref{E:P_marg}) in the minimum (top) and the fiducial (bottom) models, with the dark and light shaded regions marking the $68\%$ and $95\%$ confidence intervals, respectively. The green and blue solid curves show the PDFs of the dipole amplitudes for the pure kinematic (Eq.~\ref{E:P(d_k)_1D}) and local-source (Eq.~\ref{E:P(d_r)_1D}) terms, after marginalizing over $\sigma_r$, respectively. The blue dashed and dotted lines give the PDFs for the clustering and shot-noise terms, respectively, obtained by setting $\sigma_r^2 \approx 3C_1^{\rm clus}/(f_{\rm sky} 4\pi)$ and $\sigma_r^2 \approx 3C_1^{\rm SN}/(f_{\rm sky} 4\pi)$ in Eq.~\ref{E:P(d_r)_1D}, and marginalizing over $\sigma_r$. Colored data points show NVSS dipole measurements obtained with a 15 mJy flux threshold (see Table~\ref{T:dipole} for details). The dipole amplitude is normalized to that of the expected kinematic dipole signal. The measurements all come from previous NVSS analyses and so are drawn from the same data rather than fully independent estimates.}
\end{center}
\end{figure}

The 1D marginalized PDF of the dipole amplitude $P(d)$ (Eq.~\ref{E:P(d)} and Eq.~\ref{E:P_marg}) is displayed in Fig.~\ref{F:Pd}. In order to understand the role of the different contributions here, we show the dipole amplitude PDF for the pure kinematic term (Eq.~\ref{E:P(d_k)_1D}) and the local-source term alone (Eq.~\ref{E:P(d)} after marginalizing over $\sigma_r$). As anticipated from our earlier results, we can see that the local-source term has a comparable amplitude to the kinematic dipole but a much broader prior probability distribution. This gives the total PDF (including both local-source and kinematic contributions) a larger mean and a wider spread. The model PDF for the kinematic dipole is relatively narrowly peaked compared to the PDF of the local-source dipole; this results because the variance of the local-source dipole amplitude distribution (Eq.~\ref{E:sigma_r}) is fairly large in $\Lambda$CDM, while the kinematic contribution is reasonably well-determined even accounting for uncertainties in the NVSS source properties. Although all of the NVSS measurements exceed the peak in the model PDF, they are mostly consistent with the expected distribution given the breadth of the model PDF. An exception is the measurement from \citet[GH12; ][]{2012MNRAS.427.1994G} which favors a higher amplitude than others. The reason for this apparent discrepancy is unclear. At any rate, note that the previous NVSS dipole measurements are all from the same underlying data set: they are not statistically independent and so it is unsurprising that they appear clustered together at similar $d/\overline{d}_k$. We show the separate measurements to gauge the level of consistency across different analyses in the literature.

\begin{deluxetable*}{ccccc|c|cc|cc}[h]
\tablenum{1}
\tablecaption{\label{T:dipole} Our dipole models and NVSS dipole measurements with a 15 mJy flux threshold}
\tablewidth{0pt}
\tablehead{
\colhead{Model}  & \colhead{R.A. [deg]} &  \colhead{Dec. [deg]} & \colhead{$\theta$ [deg]} & \colhead{$d$ ($\times 10^{-2}$)}
& \colhead{} & \colhead{} & \colhead{}} 
\startdata
kinematic  & $167.942\pm0.011$  & $-6.944\pm0.005$ & ---& $0.47\pm0.03$ & & & \\ 
minimum & --- & ---  & $28^{+25}_{-18}$ & $0.61^{+0.26}_{-0.24}$ & \multicolumn{5}{c}{$\sigma$ significance} \\
fiducial & --- & ---  & $33^{+31}_{-20}$ & $0.67^{+0.31}_{-0.28}$ & \multicolumn{5}{c}{w/o (w/) kinematic-clustering correlation} \\
\hline\hline
\multicolumn{4}{c}{NVSS $S > 15$ mJy  measurements} & & kinematic & minimum & fiducial & minimum & fiducial\\
Reference & R.A. [deg] & Dec. [deg] & $\theta$ [deg] & $d$ ($\times 10^{-2}$) &  &\multicolumn{2}{c}{w/ meas. err.} & \multicolumn{2}{c}{w/o meas. err.} \\
\hline
BW02 & $148\pm29$ & $31\pm31$ & $42\pm40$ & $0.8\pm0.3$ &$1.10$ &$0.19 (0.23)$ &$0.09(0.33)$ & $0.11 (0.21)$ &$0.02 (0.39)$\\
S11 & $149\pm9$ & $15\pm9$ & $29\pm13$ & $1.3\pm0.2$ &$4.14$ &$1.03 (0.87)$ & $0.59 (0.34)$ & $1.23 (1.04)$ &$0.68 (0.40)$\\
GH12 & $117\pm20$ & $6\pm14$ & $52\pm24$ & $2.7\pm0.5$ &$4.46$ &$3.34 (3.23)$ &$2.94 (2.71)$ & $6.06 (6.07)$ &$4.62 (4.96)$\\
RS13-3D & $149\pm18$ & $-17\pm18$ & $21\pm25$ & $1.6\pm0.5$ &$2.26$ &$1.29 (1.18)$ &$1.07 (0.72)$ & $2.23 (1.93)$ &$1.49 (1.08)$\\
RS13-2D & $133\pm14$ & --- & --- & $1.5\pm0.4$ &$2.57$ & --- &--- & --- &---\\
TN16 & $151$ & $-6$ & $17$ & $0.9\pm0.4$ &$1.07$ & --- &--- & $0.41 (0.24)$ & $0.46 (0.26)$\\
S21-d & $138.90\pm12.02$ & $-2.74\pm12.11$ & $29.23\pm11.07$ & $1.6\pm0.3$ &$3.76$ &$1.64 (1.47)$ &$1.12 (0.84)$ & $2.25 (2.01)$ &$1.44 (1.17)$\\
S21-n & $156.33\pm17.80$ & $7.41\pm17.63$ & $18.44\pm15.16$ & $1.4\pm0.4$ &$2.32$ &$1.13 (0.92)$ &$0.80 (0.40)$ & $1.59 (1.32)$ &$1.05 (0.60)$\\
W23-Q32 & $145\pm13$ & $-5\pm17$ & $23\pm21$ & $1.35\pm0.30$ &$2.93$ &$1.11 (0.89)$ &$0.69 (0.42)$ & $1.39 (1.15)$ &$0.86 (0.46)$\\
W23-Q64 & $150\pm13$ & $-10\pm17$ & $18\pm21$ & $1.23\pm0.29$ &$2.62$ &$0.82 (0.63)$ &$0.52 (0.22)$ & $1.09 (0.84)$ &$0.73 (0.28)$\\
W23-P32 & $146\pm12$ & $-5\pm15$ & $22\pm19$ & $1.40\pm0.29$ &$3.20$ &$1.26 (1.02)$ &$0.81 (0.59)$ & $1.55 (1.31)$ &$0.98 (0.57)$\\
W23-P64 & $151\pm12$ & $-10\pm15$ & $17\pm19$ & $1.40\pm0.29$ &$3.13$ &$1.22 (0.97)$ &$0.81 (0.41)$ & $1.55 (1.28)$ &$1.05 (0.58)$\\
\hline\hline
\enddata
\tablecomments{BW02: \citet{2002MNRAS.329L..37B}; S11: \citet{2011ApJ...742L..23S}; GH12: \citet{2012MNRAS.427.1994G}; RS13-3D/2D: \citet{2013AA...555A.117R} linear 3D/2D estimator; TN16: \citet{2016JCAP...03..062T}; S21-d/n: \citet{2021AA...653A...9S} mask d/n; W23-Q/P 32/64: \citet{2023arXiv230515335W} quadratic/Poisson estimator with $N_{\rm side}=32/64$. The table summarizes the amplitude and direction of NVSS dipole measurements from the literature, and quantifies the level of agreement with our minimum and fiducial dipole models. It also shows the significance that would be deduced if one compares the measured dipoles with only the expected kinematic dipole signal, i.e. neglecting the local-source contributions from shot noise and clustering. The R.A. and Dec. for kinematic dipole model in the top row give the CMB dipole direction from \citet{2020AA...641A...1P}. For TN16, since they do not report the angle uncertainties, we only calculate the $\sigma$ significance of their measurement with respect to the kinematic dipole amplitude. Similarly, for the linear 2D method in RS13, which only constrains the R.A. but not the Dec. of the dipole, we also only calculate the $\sigma$ significance of their measurement with respect to the kinematic dipole amplitude. We consider cases of both including (denoted ``w/ measurement errors'') and excluding (``w/o measurement errors'') the reported measurement uncertainties. As described in Sec.~\ref{S:results}, these cases represent lower and upper limits for the inconsistency between the model and data. The $\sigma$ significance values in parentheses are calculated after including a model for the correlation between the kinematic and clustering dipoles (Sec.~\ref{S:kin_clus_corr}).}
\end{deluxetable*}

\begin{figure}[ht!]
\begin{center}
\includegraphics[width=\linewidth]{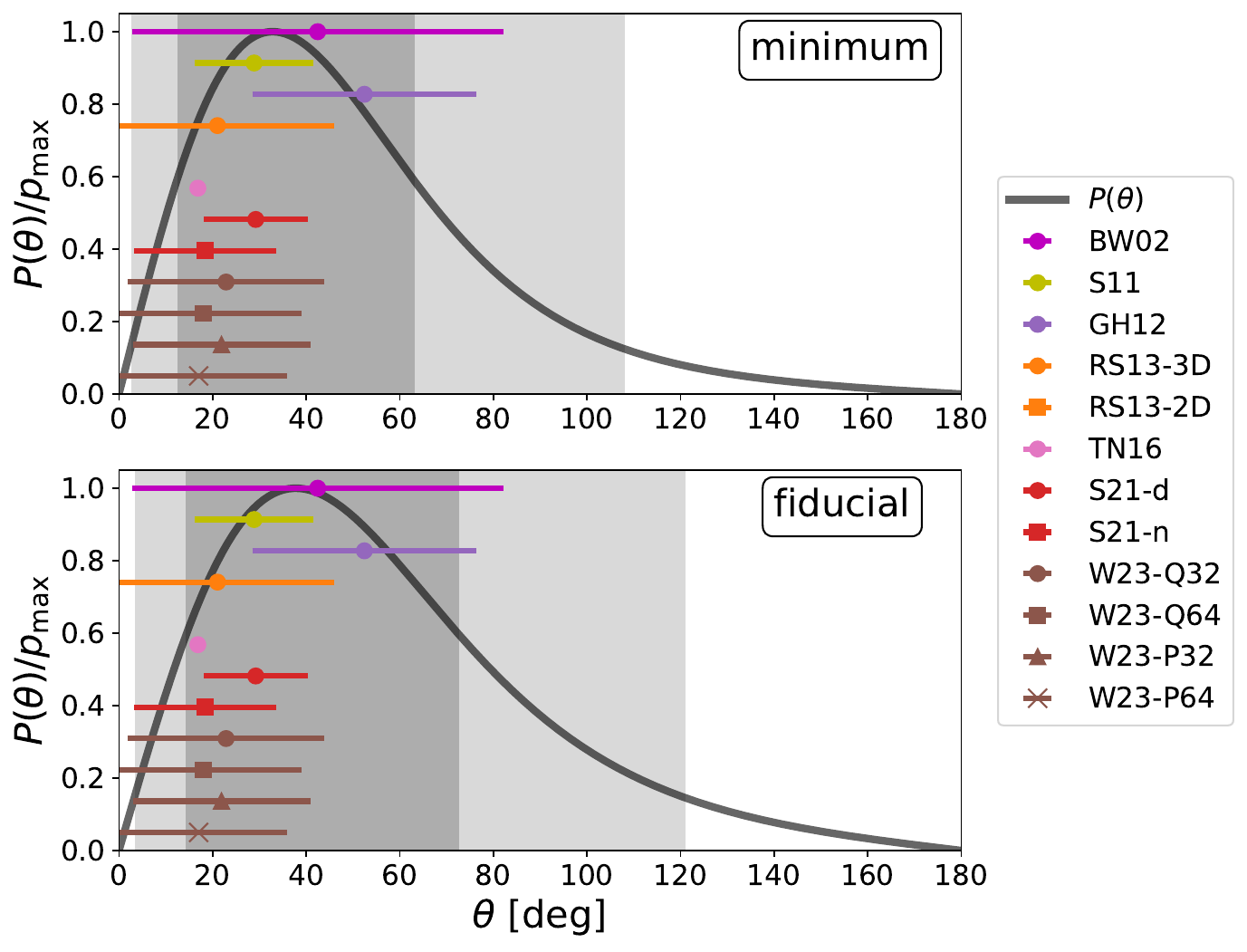}
\caption{\label{F:Pth} The 1D marginalized PDF, $P(\theta)$, for the expected NVSS dipole direction angle relative to the CMB dipole (black; Eq.~\ref{E:P(th)} and \ref{E:P_marg}) in the minimum (top) and the fidicial (bottom) models. The dark and light shaded regions mark the $68\%$ and $95\%$ confidence intervals, respectively. Colored data points show NVSS dipole measurements obtained with a 15 mJy flux threshold (see Table~\ref{T:dipole} for details).}
\end{center}
\end{figure}

Fig.~\ref{F:Pth} presents the 1D marginalized PDF for the expected dipole angle $P(\theta)$ (Eq.~\ref{E:P(th)} and Eq.~\ref{E:P_marg}) along with measurements from the literature. The dipole directions from previous NVSS measurements are all within the $1\sigma$ range of the model. As explained further below, even though the total dipole contains contributions from shot-noise and clustering, their vector sum with the kinematic contribution is expected to lie close in direction to the kinematic dipole vector, at least provided that the CMB dipole is entirely kinematic and that the LSS and CMB share the same frame. This is the case even though the model adopted here for the prior probability assumes a random angle between the kinematic and local-source dipole vectors (see Sec.~\ref{S:kin_clus_corr} for further discussion regarding the random angle assumption). The alignment arises because the kinematic term is still typically more than half of the combined signal and so the total dipole vector will largely share the direction of the pure kinematic vector. Therefore, it is not surprising that previous studies found rough agreement between the NVSS dipole direction and the CMB dipole, despite the important contribution of the local-source term to the total measured dipole. Again, this reflects the coincidence that the local-source and kinematic terms happen to be similar in magnitude. 

We can now better understand the pear-shaped contours in Fig.~\ref{F:Pdth}. These reflect the fact that high values of $d/d_k$ require both a large draw from the distribution of shot-noise plus clustering amplitudes \emph{and that the local-source dipole happens to land relatively close in direction to the kinematic one}. By contrast, at lower $d/d_k$, one can have a smaller draw from the local-source amplitude distribution and/or the local-source dipole may point further away from the CMB dipole direction. However, in all cases, the total dipole receives contributions from the kinematic term and so is unlikely to be too far in angle from the direction of the CMB dipole.

We then quantify the agreement between the NVSS dipole measurements in the literature and our PDF models in the 2D amplitude-direction plane. One subtlety in properly assessing the statistical significance here is that some portion of the reported measurement errors arise from features that are already incorporated in our model PDFs, and so there is a danger of double-counting portions of the error budget. For example, in some studies the error bars are estimated from the spread across simulated realizations of randomly-distributed sources with a given kinematic dipole imposed. In this case, shot-noise is accounted for in the error budget, but not the clustering component. On the other hand, there are potentially important measurement errors that are not accounted for in our model PDFs, such as flux measurement uncertainties. Since it is not always straightforward to determine which sources of uncertainty are included in the measurement errors reported in the literature, and their relative contributions, we consider two limiting cases that establish lower and upper bounds on the level of consistency with our models.

First, we approximate the measurement errors as entirely independent of the width of our model PDF. That is, in this case we suppose the measurement errors describe entirely different sources of uncertainty than captured by our model PDF. Here, we take the measurement errors at face value, and suppose there is no double-counting with the model uncertainties. This establishes an upper limit for the level of consistency between the model and data (i.e. it places a lower bound on the $\sigma$ level).


In the second case, we consider an approximation where the reported error bars have already taken into account {\em all} of the factors incorporated within our PDF. Additionally, we assume that other measurement uncertainties are negligible. This establishes a lower limit for the level of agreement between model and data (i.e. it yields an upper bound on the $\sigma$ level). 
In this context, the error reported from previous measurements should not be included in to prevent double-counting. 
The level of consistency between each measurement and our minimum and fiducial models, under these two limiting scenarios, is presented in Table~\ref{T:dipole} in terms of their $\sigma$ significance levels listed under the ``w/''  (upper bound) and ``w/o measurement error'' (lower bound) columns, respectively. In addition, we also list $\sigma$ level after incorporating the clustering and kinematic dipole correlation (see Sec.~\ref{S:kin_clus_corr} for details) in each corresponding case.

Our analysis indicates that the previous measurements (with the exception of GH12) deviate from the minimum/fiducial model by no more than $\sim 2.25\sigma/1.49\sigma$ in terms of statistical significance, even in the scenario of maximum inconsistency, where we disregard any contribution from the measurement error budget in the computation. We also emphasize that after properly accounting for the kinematic and clustering dipole correlation, most of the previous measurements are even consistent with the CMB dipole at $\lesssim 1\sigma$, at least in our fiducial case with measurement errors included. These results suggest that, within the current statistical uncertainties, the NVSS measurements are consistent with the expectations from the CMB dipole observations. Although our imperfect knowledge of the NVSS redshift distribution and clustering impacts the details -- as reflected in the difference between our fiducial and minimum models -- the general conclusion of consistency with the CMB expectations is insensitive to the difference between these two cases. 

Our conclusions here differ from the majority of previous NVSS dipole studies, which claimed discrepancies with the CMB expectations. This reflects two key differences between our analysis and previous work. First, many previous studies determined the statistical significance based on the dipole amplitude alone, without considering the joint distribution of the amplitude and angle as we did in our analysis. The exception is \citet{2022ApJ...937L..31S}, who found a $2.6\sigma$ significance for the NVSS dipole compared to expectations using the joint PDF $P(d,\theta)$. However, their model does not account for local-source contributions. Nevertheless, using the joint PDF of dipole amplitude and angle is not the main cause of our differences with previous work since previous measurements are generally consistent in direction with the CMB dipole.

Instead, the key difference is that most studies claiming discrepancies compared their NVSS dipole measurements with the expected kinematic dipole alone, neglecting the clustering and/or shot-noise terms when modeling the expected dipole. Note that the shot noise and clustering impact both the expected values of the model PDFs, as well as the width of the distribution in amplitude and angle. A proper assessment requires accounting for the role of shot noise and clustering in changing the position of the peak in the model PDF as well as their impact on the width of the distribution. For comparison, we also calculate the $\sigma$-significance levels of previous measurements using our $P(d_k)$ (green curves in Fig.~\ref{F:Pd}). In this case, we observe that many of the measurements are inconsistent with the CMB dipole at a significance level of $\sim3-4\sigma$, similar to several claims in the literature.By contrast, our analysis incorporates a detailed model for all dipole contributions and derives the PDF of the total dipole, with the local-source contributions playing an important role.

\section{Discussion}\label{S:discussion}
\subsection{Additional Modeling and Data Processing Uncertainties}\label{S:discussion_systematics}
Our analysis thus far has neglected a few potentially important ingredients, which we discuss here. First, we discuss possible correlations between the kinematic and local-source dipoles (Sec.~\ref{S:kin_clus_corr}). Next, we quantify the expected bulk motion of the NVSS sources relative to the CMB frame (Sec.~\ref{S:bulk_flow}), finding it to be negligibly small. Sec.~\ref{S:discussion_srcselection_masking} discusses further potential systematic uncertainties related to processing the NVSS data. In Sec.~\ref{S:partial_sky} we expand on the issue of partial sky coverage, while Sec.~\ref{S:future} considers future dipole measurement prospects.

\subsection{Correlation between the Kinematic and Local-source Dipoles}\label{S:kin_clus_corr}
As briefly mentioned earlier, there may be a correlation between the kinematic and local-source dipoles, yet this has been neglected thus far. This correlation results because the clustering of the NVSS sources traces mass inhomogeneities, which also source our peculiar motion and the kinematic dipole. These will not perfectly correlate because the clustering dipole and our peculiar velocity have different wavenumber/redshift weightings -- that is, the clustering and peculiar motion terms may be expressed as integrals over wavenumber and redshift that only partly overlap (see Appendix~\ref{A:kin_clus_corr} for details) -- but this could nevertheless be important. If this correlation is non-negligible and positive, it will tend to boost the expected total dipole amplitude and the alignment with the CMB dipole. On the other hand, shot noise and the motion of the Sun relative to the the Local Group add random contributions to the local-source and kinematic dipoles, respectively, and will tend to reduce this correlation. This correlation has also been accounted for in the quasar dipole analysis of \citet{2023MNRAS.525..231D}.

To quantify the impact of this correlation, we start by calculating the expected correlation coefficient between $\mathbf{d}_k$ and $\mathbf{d}_r$:
\begin{equation}\label{E:vd_corr_coeff}
r_{\rm vd} = \frac{\langle\mathbf{d}_k\cdot\mathbf{d}_r\rangle}{\sqrt{\langle\mathbf{d}_k\cdot\mathbf{d}_k\rangle\langle\mathbf{d}_r\cdot\mathbf{d}_r\rangle}}=\frac{\langle\mathbf{v}_k\cdot\mathbf{d}_r\rangle}{\sqrt{\langle\mathbf{v}_k\cdot\mathbf{v}_k\rangle\langle\mathbf{d}_r\cdot\mathbf{d}_r\rangle}},
\end{equation}
where the velocity, $\mathbf{v}_k$, is proportional to the kinematic dipole, $\mathbf{d}_k$.

As usual, the local-source dipole can be decomposed into the clustering term ($\mathbf{d}_{\rm clus}$) and the shot-noise term ($\mathbf{d}_{\rm SN}$):
\begin{equation}
\mathbf{d}_r = \mathbf{d}_{\rm clus} + \mathbf{d}_{\rm SN},
\end{equation}
where only the clustering term is correlated with the kinematic one. The peculiar motion consists of a smooth component ($\mathbf{v}_R$) and a stochastic term ($\mathbf{v}_s$):
\begin{equation}
\mathbf{v}_k = \mathbf{v}_{R} + \mathbf{v}_s.
\end{equation}
The smooth component represents the coherent motion of our Local Group, as induced by the LSS, and is therefore partly correlated with $\mathbf{d}_{\rm clus}$. On the other hand, the stochastic component accounts for the additional scatter due to random motions in the local gravitational potential. That is, the kinematic dipole reflects the motion of the Sun relative to the CMB frame, which partly owes to the bulk velocity of the Local Group, as sourced by surrounding mass inhomogeneities, but also depends on the Sun's velocity relative to the Local Group. This, in turn, depends on the motion of the Sun relative to the local standard of rest, which itself orbits around the Galactic Center, while the Milky Way moves with respect to the Local Group. It is only the bulk motion of the Local Group that is sourced by the LSS and so these additional components can be thought of as stochastic contributions for present purposes.  Hence, the three inner products in Eq.~\ref{E:vd_corr_coeff} can be expressed as follows:
\begin{equation}
\begin{split}
\langle\mathbf{v}_k\cdot\mathbf{d}_r\rangle =& \langle\mathbf{v}_R\cdot\mathbf{d}_{\rm clus}\rangle\\
\langle\mathbf{d}_r\cdot\mathbf{d}_r\rangle =& \langle\mathbf{d}_{\rm clus}\cdot\mathbf{d}_{\rm clus}\rangle+\langle\mathbf{d}_{\rm SN}\cdot\mathbf{d}_{\rm SN}\rangle\\
=& \frac{9}{4\pi}\left(C_1^{\rm clus} + C_1^{\rm SN}\right)\\
\langle\mathbf{v}_k\cdot\mathbf{v}_k\rangle =& \langle\mathbf{v}_R\cdot\mathbf{v}_R\rangle+\langle\mathbf{v}_s\cdot\mathbf{v}_s\rangle=\sigma_R^2+\sigma_s^2,
\end{split}
\end{equation}
where $\sigma^2_R$ and $\sigma^2_s$ are the variance of $\mathbf{v}_R$ and $\mathbf{v}_s$, respectively. We have models for the clustering and shot-noise dipole variances, $\langle\mathbf{d}_{\rm clus}\cdot\mathbf{d}_{\rm clus}\rangle$ and $\langle\mathbf{d}_{\rm SN}\cdot\mathbf{d}_{\rm SN}\rangle$, derived in Sec.~\ref{S:clus_dipole} and \ref{S:SN_dipole}, respectively. The coherent component, $\mathbf{v}_R$, can be calculated using the continuity equation and linear perturbation theory:
\begin{equation}\label{E:vR_3D}
\mathbf{v}_R = -if_\Omega H_0\int \frac{d^3\mathbf{k}}{(2\pi)^3}\frac{\mathbf{k}}{k^2}W_v(kR)\delta_m^{\rm 3D}(\mathbf{k}),
\end{equation} 
where $H_0$ is the Hubble constant, $f_\Omega$ is the growth rate at $z=0$, $\delta_m^{\rm 3D}(\mathbf{k})$ is the 3D matter density field in Fourier space, while $W_v(kR)=3j_1(kR)/(kR)$ is the Fourier transform of a spherical top-hat window with a radius $R$. The window function here describes an average over the Local Group scale ($R \sim$ Mpc). With this expression along with our clustering model, we can derive formulas for $\langle\mathbf{v}_R\cdot\mathbf{d}_{\rm clus}\rangle$ and $\langle\mathbf{v}_R\cdot\mathbf{v}_R\rangle$ (Eq.~\ref{E:vR_dclus_corr} and \ref{E:vR_vR_corr}; see Appendix~\ref{A:kin_clus_corr} for detailed derivations). For the rms stochastic velocity contribution, a plausible estimate is that the ensemble-averaged ratio between the random velocity and the coherent bulk flow match that in our particular solar system/Local Group realization. In this case, according to \citet{2020AA...641A...1P}, the relative motion of the solar system and the center of the Local Group is $v_{\rm Sun-LG}=299\pm15$ km s$^{-1}$, while the speed of the Local Group relative to the CMB is $v_{\rm LG-CMB}=620\pm15$ km s$^{-1}$. Therefore, we assume $\sigma_s^2/\sigma_R^2=299^2/620^2=0.23$, which results in a $\sim10\%$ decrease in the correlation coefficient, $r_{\rm vd}$, as compared to the value without the stochastic velocity term. 

Assuming a velocity window of radius $R=1$ Mpc $h^{-1}$, representing the scale of the Local Group, our minimum and fiducial clustering models yield correlation coefficients of $r_{\rm vd}=0.21$ and $r_{\rm vd}=0.45$, respectively. The values of $r_{\rm vd}$ are insensitive to the choice of velocity window radius, $R$. Interestingly, these correlations appear non-negligible for the NVSS sample. We incorporate these correlations into our model for the PDF of the total measured dipole (Appendix~\ref{S:pdf_kin_clus_included}). The corresponding results for the joint and marginalized model PDFs are shown in  Fig.~\ref{F:Pdth_vdcorr}, \ref{F:Pd_vdcorr}, and \ref{F:Pth_vdcorr}. As anticipated, including this correlation results in an increased dipole amplitude and a shift in the offset angle relative to the CMB dipole towards smaller values, further enhancing the consistency between our model predictions and the measurements (Table~\ref{T:dipole}). This is because most of the previous measurements tend to have slightly higher amplitude and smaller offset angle compared to the peak in the probability distribution in our baseline models, where we neglect the correlation between the kinematic and local-source dipole.

\subsection{Bulk Flow Velocity}\label{S:bulk_flow}
In principle, the NVSS sources might not exactly share the CMB rest frame. That is, the NVSS objects might themselves have a net bulk flow relative to the more distant CMB reference frame, as sourced by density inhomogeneities far away, even under the standard $\Lambda$CDM cosmology. This is, however, expected to be small given that the NVSS source distribution spans a large cosmological volume. 

The bulk flow is expected to have a random direction, and its amplitude is determined by the vector sum of the peculiar velocities of the NVSS sources. We can calculate the variance of the bulk flow velocity using linear perturbation theory (see Appendix~\ref{A:bulk_flow}). The bulk flow velocity variance, $\langle\mathbf{v}^2_{\rm bulk}\rangle$ (Eq.~\ref{E:v_bulk}), depends on the window function, $\frac{dN}{dz}$, of the NVSS sources. Although our models inferred from \textit{Tomographer} only constrain $b(z)\frac{dN}{dz}$, we can still place an upper limit on $\langle\mathbf{v}^2_{\rm bulk}\rangle$ by assuming $b(z)=1$. By using the mean value of $b(z)\frac{dN}{dz}$ from our models for $\frac{dN}{dz}$ in Eq.~\ref{E:v_bulk}, we find upper bounds on the rms bulk flow velocities of 2.46 km s$^{-1}$ and 4.82 km s$^{-1}$ for the minimum and fiducial models, respectively. These values are only $\sim 1\%$ of the motion inferred from the CMB dipole \citep[$v_{\rm kin} = 369.82\pm0.11$ km s$^{-1}$; ][]{2020AA...641A...1P}. Therefore, we conclude that the effects of such bulk flows are negligible for the NVSS sources.

\subsection{Source Selection and Masking}\label{S:discussion_srcselection_masking}
One potential concern arises from the differences in our source selection and masking criteria compared to previous NVSS measurements. In establishing our clustering and shot-noise models, our analysis focuses solely on smaller-scale NVSS clustering measurements. This reduces the susceptibility to several large-scale systematic effects associated with the dipole measurements themselves. Furthermore, we have conducted tests to ensure the stability of our clustering and shot-noise models under variations in masking radius, flux threshold, and the inclusion of additional regions with high source density. Therefore, our specific choice of source selection and masking scheme is unlikely to introduce significant systematic errors into our modeling.

Another potential systematic in our modeling is related to the treatment of the clustering dipole at $z<0.01$. We assume that NVSS sources below this redshift have been completely removed, which may lead to an underestimate of the clustering dipole. Previous studies employ different strategies to remove nearby sources and extended regions with high source density to partly mitigate the systematics from the nearby sources. However, quantifying the impact of remaining nearby sources in the data is challenging. If all nearby sources, at $10^{-3} < z < 0.01$, remain and follow the redshift distribution calibrated by our cross-matching, we estimate that they will contribute an additional $~20\%$ to the clustering dipole signal (Sec.~\ref{S:clus_dipole}). Therefore, it is possible that the clustering dipole in previous measurements is higher than our model by $\lesssim20\%$. We emphasize that our minimum model sets a lower bound on the clustering dipole from $z<0.01$ sources by assuming that any source from $z<10^{-3}$ has been completely removed (as is assumed in our fiducial scenario too), by adopting a clustering bias of unity at $10^{-3} < z < 0.01$, and by supposing that all NVSS sources in this range have successfully cross-matched counterparts in 2MRS. 
 Even with these conservative assumptions, our calculations show that this minimal clustering contribution, along with shot-noise, alleviates the previously claimed NVSS dipole measurement discrepancies. Therefore, taking into account the incompleteness of our low-redshift cross-matching procedure, and the clustering bias of such sources, makes our prediction of the total dipole amplitude still more consistent with previous measurements (as is the case in our fiducial model).

 \subsection{Partial Sky Correction}\label{S:partial_sky}
A caveat of our analysis regards the treatment of partial sky coverage. The incomplete sky coverage of NVSS, and the additional masking employed, lead to mode-mixing effects. Further, the shape of the model PDF will deviate from the idealized full-sky limit assumed in our derivations: for example, our model PDF assumes perfect isotropy while this will be violated in a realistic survey. Instead, here we adopted the simple approximation that the form of the PDF follows the full-sky case, but with enhanced shot-noise plus clustering variance scaling as $1/f_{\rm sky}$, akin to the CMB angular power spectrum uncertainties \citep{1995PhRvD..52.4307K}. This accounts for leakage (into the dipole) from higher-$\ell$ shot-noise and clustering power under the approximation that these contributions follow a flat white-noise power spectrum (see also \citealt{2021JCAP...11..009N}).  Using our value of $f_{\rm sky} = 0.63$ (the precise sky fraction varies slightly from study to study), the rms fluctuations, $\sigma_r$ (Eq.~\ref{E:sigma_r}), increase by approximately $\sim 25\%$.

In practice, systematic errors in the survey maps may also lead to spurious power in higher-order ($\ell \geq 2$) multipoles that then leak into the dipole estimates in the cut sky. In addition to the dependence on any remaining systematics in the data, the precise partial-sky effects will depend on the details of the masks employed in each study and the particular dipole estimator used in the analysis. We hence do not consider the cut-sky effects in more detail here. In future work, it will be valuable to construct mock full-sky simulations including random realizations of the shot-noise and clustering (with appropriate angular power spectra, as calibrated in this work), along with draws from the kinematic dipole distribution. Further, one can include mock systematic fluctuations with various angular power spectra to investigate their potential impact. One can then apply arbitrary masks to the simulated full-sky maps, and determine the PDF of the resulting cut-sky dipole estimates after averaging over many random realizations. In the full-sky, systematics-free limit, this will recover our analytic PDF while the masked simulations can be used to refine our present estimates for the model PDFs with incomplete sky coverage.

\subsection{Future Prospects}\label{S:future}
Our analysis demonstrates that the local-source terms likely contribute significantly to the total NVSS dipole signal. This poses a challenge for using NVSS to provide better measurements of the kinematic dipole and sharper tests of the Cosmological Principle. Future surveys, however, can help by selecting only objects at high redshift to mitigate the clustering dipole contributions, and by achieving higher source densities to reduce shot-noise in the measurements.

Recently, a few studies have conducted dipole measurements using the CatWISE quasar samples \citep{2021ApJ...908L..51S,2023MNRAS.525..231D}. The redshift distribution of this sample is favorable, with negligible numbers of sources in the nearby universe, and this should help minimize local clustering contributions. However, the source number densities in this catalog are comparable to those in NVSS and so this leads to similar shot-noise levels which hence impact the dipole estimates from this data set.

Upcoming cosmological surveys are expected to provide improved data sets for the dipole test using high-redshift LSS tracers. First, \citet{2019MNRAS.486.1350B} predicts that the radio dipole amplitude can be determined to $\sim10\%$ precision with the forthcoming SKA \citep{2020PASA...37....7S} survey. \citet{2022MNRAS.515L..11K} explore the potential of probing the dipole using near-infrared photometric sources from the Euclid mission \citep{2018LRR....21....2A} and the Roman Space Telescope \citep{2015arXiv150303757S}. In addition, SPHEREx \citep{2014arXiv1412.4872D,2018arXiv180505489D}\footnote{\url{http://spherex.caltech.edu}} is designed to perform an all-sky near-infrared spectroscopic galaxy survey with unprecedented depth and well-controlled large-scale systematics. SPHEREx will be ideal for probing the dipole with minimal contamination from local sources. By selecting objects at higher redshifts ($z\gtrsim 0.5$), the contribution from the clustering term can be significantly reduced (see Fig.~\ref{F:kz_kernel}). Additionally, the expected source number density from SPHEREx will suppress shot noise to a negligible level. Furthermore, the low-resolution spectra for each source in the SPHEREx catalog will provide valuable information for accurately modeling the kinematic dipole from these objects, as the kinematic dipole amplitude depends on the spectral shape of each source and the number counts as a function of specific flux (see Eq.~\ref{E:dkin}). This improved modeling will enable a more robust test of the Cosmological Principle.

\section{Conclusion}\label{S:conclusion}
Several analyses of the dipole moment in the angular distribution of NVSS sources have found disagreement with the expectations from the well-determined CMB dipole. In this work, we built a comprehensive model for the prior probability of the dipole measured from NVSS.

We modeled the two components of the local-source dipole, clustering and shot-noise, using the NVSS data itself. To determine the clustering signal, we employed the clustering redshift technique to infer the redshift kernel for the clustering dipole at $z>0.1$. We also used a cross-matching approach between the NVSS and 2MRS catalogs to infer the redshift distribution of more nearby NVSS sources. With these two methods, we construct a ``minimum'' and a ``fiducial'' model for the clustering dipole. Combining this with our models for the kinematic dipole and the shot-noise, we derive the PDF of the expected NVSS dipole, taking into account the statistics of both the kinematic and local-source dipoles.

Our models identify an interesting coincidence: the NVSS local-source dipole happens to be comparable in amplitude to the kinematic dipole in spite of the disparate origins of these two contributions. This then helps reconcile previous NVSS dipole estimates with the CMB expectations. In particular, the expected total dipole is higher after considering the combination of these two terms, contrary to the assumptions made in many previous works where measurements were compared only to the kinematic dipole term. Quantitatively, we find that most of the previous dipole measurements are consistent with the CMB dipole, under the kinematic origin hypothesis and $\Lambda$CDM, at better than $\sim 2 \sigma$, with the precise level of agreement depending on the particular dipole measurement (see Table~\ref{T:dipole}). At present, we find that more exotic explanations for the CMB and/or NVSS dipoles are not required by these data. We note, however, that the LSS dipole has also been measured with other radio catalogs (see \citealt{2021AA...653A...9S} for a list of previous measurements), and some analyses find even higher dipole amplitudes than in NVSS. Further tests for systematic errors in these catalogs, as well as investigations of their local-source contributions are needed to reliably cross-check the consistency of these measurements with CMB expectations.

In addition, recent work by \citet {2021ApJ...908L..51S} and \citet{2023MNRAS.525..231D} find, respectively, a $4.9\sigma$ and a $5.7\sigma$ discrepancy between the dipole in the CatWISE quasar sample and the CMB dipole expectations. The latter analysis carefully accounted for the local-source contributions. More detailed investigations of potential systematic effects in the measurements, additional modeling of the CatWISE samples, and other independent measurements with LSS tracers are essential further tests here. For example, the recent quasar catalog in \citet{2023arXiv230617749S}, constructed from a joint analysis with Gaia and unWISE, may be helpful in this regard.

Upcoming cosmological surveys, including SKA, Euclid, Roman, and SPHEREx promise to provide much better datasets for performing dipole measurements. These surveys will have higher source densities and the ability to isolate samples of purely high redshift objects, which will significantly reduce both the shot-noise and clustering contributions to the dipole signal, while spectroscopic information will help in determining the expected discrete source kinematic dipole. Therefore, these future surveys should provide more powerful 
tests of the Cosmological Principle and further cross-checks regarding the kinematic origin of the CMB dipole. 

\acknowledgments
We would like to thank the anonymous referee for valuable comments that improved the manuscript. We are grateful to them, Chris Anderson, Jordan Mirocha, Guillem Dom{\'e}nech, Dragan Huterer, Eoin \'O Colg\'ain, and Nathan Secrest for constructive comments on a draft manuscript, as well as, Olivier Dor{\'e}, Giulio Fabbian, Yen-Ting Lin, Ue-Li Pen, Andrew Romero-Wolf, Douglas Scott, Mike Seiffert, Wei-Hao Wang, Abby Williams for helpful discussions throughout this work. We thank Yi-Kuan Chiang for advice regarding the use of \emph{Tomographer} and Judah Luberto for early collaboration on topics related to this paper. AL is grateful to the JPL Visitor Program for hospitality and support during a sabbatical visit to JPL. Y.-T.C. acknowledges support by NASA ROSES grant 18-ADAP18-0192. T.-C.C. acknowledges support by NASA ROSES grants 17-ADAP17-0234 and 21-ADAP21-0122. Y.-T.C. and T-C.C. were also supported in part by the JPL 7X formulation fund. Part of this work was done at Jet Propulsion Laboratory, California Institute of Technology, under a contract with the National Aeronautics and Space Administration. 
\software{
astropy \citep{2013A&A...558A..33A,2018AJ....156..123A}, HEALPix \citep{2005ApJ...622..759G}, healpy \citep{2019JOSS....4.1298Z},
NaMaster \citep{2019MNRAS.484.4127A}}

\vspace{3pt}

\appendix

\section{PDF of the Dipole Signal}\label{A:formalism}
Here we provide a detailed derivation of the model PDF adopted in our analysis. As discussed in the manuscript and Sec.~\ref{S:partial_sky} our model only approximately accounts for partial sky coverage effects. The measured dipole, $\mathbf{d}$, is the vector sum of the kinematic dipole, $\mathbf{d}_k$, and the local-source dipole, $\mathbf{d}_r$ (which includes both clustering and shot-noise contributions),
\begin{equation}
\mathbf{d} = \mathbf{d}_k+ \mathbf{d}_r = \mathbf{d}_k + \mathbf{d}_{\rm clus}+ \mathbf{d}_{\rm SN}.
\end{equation}
Defining $\mathbf{\hat{z}}$ as the direction of the kinematic dipole and $d_k$ as the kinematic dipole amplitude, we can express $\mathbf{d}_k=d_{k,x} \mathbf{\hat{x}} + d_{k,y} \mathbf{\hat{y}} + d_{k,z} \mathbf{\hat{z}} = d_k \mathbf{\hat{z}}$. The PDF the vector $\mathbf{d}_k$ is modeled as a Gaussian distribution with mean $\overline{d}_k$ and variance $\sigma_k^2$:
\begin{equation}\label{E:P(d_k)_3D}
P(\mathbf{d}_k) = \delta^D(d_{k,x})\delta^D(d_{k,y})\frac{1}{\sqrt{2\pi\sigma_k^2}}e^{-\frac{(d_{k,z}-\overline{d}_k)^2}{2\sigma_k^2}}.
\end{equation}

Ignoring for now the correlation between the kinematic and local-source dipoles (see Sec.~\ref{S:kin_clus_corr}, Appendix~\ref{A:kin_clus_corr}), the PDF of the local-source dipole vector $\mathbf{d}_r$, for a given rms $\sigma_r$, follows a zero-mean 3D Gaussian distribution:
\begin{equation}\label{E:P(d_r)_3D}
P(\mathbf{d}_r|\sigma_r) = \frac{1}{(2\pi\sigma_r^2)^{3/2}}e^{-\frac{d_{r,x}^2+d_{r,y}^2+d_{r,z}^2}{2\sigma_r^2}},
\end{equation}
where $\sigma_r^2$ is proportional to the angular power spectrum,
\begin{equation}\label{E:sigma_r_appendix}
\sigma^2_r \approx \frac{1}{f_{\rm sky}}\frac{3}{4\pi}\left(C_1^{\rm clus} + C_1^{\rm SN}\right).
\end{equation}
We construct the PDF of the clustering power spectrum, $C_1^{\rm clus}$, accounting for redshift uncertainties (Sec.\ref{S:clus_dipole}).  From this, we can derive the PDF of $\sigma_r^2$ using the chain rule:
\begin{equation}\label{E:P(sigmar)}
P(\sigma_r) = \frac{\partial C_1^{\rm clus}}{\partial \sigma_r}P(C_1^{\rm clus})=\frac{8\pi\sigma_r}{3}P(C_1^{\rm clus}).
\end{equation}

The 1D PDFs of the kinematic and local-source dipole amplitudes can be derived from Eq.~\ref{E:P(d_k)_3D} and \ref{E:P(d_r)_3D}:
\begin{equation}
P(d_k) = \frac{1}{\sqrt{2\pi\sigma_k^2}}e^{-\frac{(d_k-\overline{d}_k)^2}{2\sigma_k^2}},
\end{equation}
\begin{equation}\label{E:P(d_r)_1D}
P(d_r|\sigma_r) = \frac{4\pi d_r^2}{(2\pi\sigma_r^2)^{3/2}}e^{-\frac{d_r^2}{2\sigma_r^2}}.
\end{equation}

The PDF of the total measured dipole $\mathbf{d}$ for a given $\sigma_r$ is given by
\begin{equation}\label{E:P(d_3D)}
\begin{split}
&P(\mathbf{d}|\sigma_r) = \int d^3 \mathbf{d}_k P(\mathbf{d}_k) \int d^3 \mathbf{d}_r P(\mathbf{d}|\mathbf{d}_k, \mathbf{d}_r) P(\mathbf{d}_r|\mathbf{d}_k, \sigma_r)\\
&=\int dd_k \frac{1}{\sqrt{2\pi\sigma_k^2}}e^{-\frac{(d_{k}-\overline{d}_k)^2}{2\sigma_k^2}} \\
&\cdot\int d^3 \mathbf{d}_r \delta^D(\mathbf{d}-(\mathbf{d}_k+\mathbf{d}_r))\frac{1}{(2\pi\sigma_r^2)^{3/2}}e^{-\frac{d_{r,x}^2+d_{r,y}^2+d_{r,z}^2}{2\sigma_r^2}}\\
&=\int dd_k \frac{1}{\sqrt{2\pi\sigma_k^2}}e^{-\frac{(d_{k}-\overline{d}_k)^2}{2\sigma_k^2}}\frac{1}{(2\pi\sigma_r^2)^{3/2}}e^{-\frac{d_x^2+d_y^2+(d_z-d_k)^2}{2\sigma_r^2}}\\
&=\frac{1}{2\pi\sigma_r^2\sqrt{2\pi(\sigma_r^2+\sigma_k^2)}}e^{-\frac{d_x^2+d_y^2}{2\sigma_r^2}}e^{-\frac{(d_z-\overline{d}_k)^2}{2(\sigma_r^2+\sigma_k^2)}}.
\end{split}
\end{equation}
Note that the assumption of uncorrelated kinematic and clustering dipoles leads to the relation $P(\mathbf{d}_r|\mathbf{d}_k, \sigma_r)=P(\mathbf{d}_r|\sigma_r)$.

To directly compare our model with measurements, we express $\mathbf{d}$ in terms of the amplitude, $d\equiv\sqrt{d_x^2+d_y^2+d_z^2}$, and the angle, $\theta$, between $\mathbf{d}$ and $\mathbf{d}_k$ (which lies in the $\mathbf{\hat{z}}$ direction). By defining $\mu\equiv {\rm cos}\,\theta=d_z/d$, $\kappa\equiv\sigma_k/\sigma_r$, and noting the relation $d^3\mathbf{d}=2\pi d^2\, dd \,d\mu$, we obtain
\begin{equation}
\begin{split}
P&(d,\mu|\sigma_r)=2\pi d^2 P(\mathbf{d}|\sigma_r)\\
&=\frac{1}{\sqrt{2\pi(\sigma_r^2+\sigma_k^2)}}\frac{d^2}{\sigma_r^2}e^{-\frac{(1+(1-\mu^2)\frac{\sigma_k^2}{\sigma_r^2})d^2-2\overline{d}_k\mu d+ \overline{d}_k^2}{2(\sigma_r^2+\sigma_k^2)}}\\
&=\frac{1}{\sqrt{2\pi\sigma_r^2(1+\kappa^2)}}\frac{d^2}{\sigma_r^2}e^{-\frac{(1+(1-\mu^2)\kappa^2)d^2-2\overline{d}_k\mu d+ \overline{d}_k^2}{2\sigma_r^2(1+\kappa^2)}},
\end{split}
\end{equation}
and
\begin{equation}
\begin{split}
P&(d,\theta|\sigma_r)={\rm sin}\,\theta P(d,\theta|\sigma_r)\\
&=\frac{{\rm sin}\,\theta}{\sqrt{2\pi\sigma_r^2(1+\kappa^2)}}\frac{d^2}{\sigma_r^2}e^{-\frac{(1+(1-\mu^2)\kappa^2)d^2-2\overline{d}_k\mu d+ \overline{d}_k^2}{2\sigma_r^2(1+\kappa^2)}}.
\end{split}
\end{equation}

The marginalized distribution for the PDF of $d$ is given by
\begin{equation}\label{E:P(d)}
\begin{split}
&P(d|\sigma_r)=\int_{-1}^{1}d\mu P(d,\mu|\sigma_r)\\
&=\frac{1}{\sqrt{2\pi\sigma_r^2(1+\kappa^2)}}\frac{d^2}{\sigma_r^2}e^{-\frac{(1+\kappa^2)d^2+ \overline{d}_k^2}{2\sigma_r^2(1+\kappa^2)}} \int_{-1}^{1}d\mu e^{\frac{\kappa^2d^2\mu^2+2\overline{d}_k d\mu}{2\sigma_r^2(1+\kappa^2)}}\\
&=\frac{d}{2\kappa\sigma_r^2}e^{-\frac{d^2+(\overline{d}_k/\kappa)^2}{2\sigma_r^2}}\left[ {\rm erfi\left(\frac{\kappa d + (\overline{d}_k/\kappa)}{\sqrt{2\sigma_r^2(1+\kappa^2)}}\right)}\right.\\
&\,\,\,\,\,\,\,\,\,\,\,\,\,\,\,\,\,\, - \left. {\rm erfi\left(\frac{-\kappa d + (\overline{d}_k/\kappa)}{\sqrt{2\sigma_r^2(1+\kappa^2)}}\right)} \right],
\end{split}
\end{equation}
where ``erfi'' is the imaginary error function. In the limit of $\sigma_k\ll\sigma_r$ and thus $\kappa\rightarrow 0$, it is useful to note that:
\begin{equation}\label{E:P(d)_kappa0}
P(d|\sigma_r)=\frac{2}{\sqrt{2\pi\sigma_r^2}}\frac{d}{\overline{d}_k}e^{-\frac{d^2+\overline{d}_k^2}{2\sigma_r^2}}{\rm sinh}\left(\frac{d\overline{d}_k}{\sigma_r^2}\right).
\end{equation}
\cite{2021JCAP...11..009N} also considered the total dipole from the kinematic and local-source terms and derived expressions for the mean and variance of $d$ given $\sigma_k$ and $\sigma_r$ (see Eq. 2.11 in \cite{2021JCAP...11..009N}). Upon checking, we found that the first and second moments of our PDF in Eq.~\ref{E:P(d)_kappa0} are equivalent to the expressions in Eq. 2.11 of \cite{2021JCAP...11..009N}.

Similarly, we can also derive the marginalized distributions for the PDFs of $\mu$ and $\theta$:
\begin{equation}\label{E:P(th)}
\begin{split}
P&(\mu|\sigma_r)=\int_{0}^{\infty}dd P(d,\mu|\sigma_r)\\
&=\frac{1}{\sigma_r^2\sqrt{2\pi\sigma_r^2(1+\kappa^2)}}e^{-\frac{\overline{d}_k^2}{2\sigma_r^2(1+\kappa^2)}}\cdot\mathcal{I},
\end{split}
\end{equation}
\begin{equation}
P(\theta|\sigma_r) = {\rm sin}\,\theta P(\mu|\sigma_r)
\end{equation}
where
\begin{equation}
\begin{split}
&\mathcal{I} = \int_0^{\infty}dd\,d^2e^{-\frac{(1+(1-\mu^2)\kappa^2)d^2 - 2\overline{d}_k\mu d}{2\sigma_r^2(1+\kappa^2)}}\\
&= \frac{bc}{a^2}+\frac{\sqrt{2\pi c}(ac+b^2)}{2a^2\sqrt{a}}e^{\frac{b^2}{2ac}}\left[{\rm erf}\left(\frac{b}{\sqrt{2ac}}\right)+1\right]\\
&a=1+(1-\mu^2)\kappa^2\\
&b=\overline{d}_k\mu\\
&c=\sigma_r^2(1+\kappa^2).
\end{split}
\end{equation}
In the limit of $\kappa\rightarrow 0$, this expression reduces to
\begin{equation}
\begin{split}
P&(\mu|\sigma_r)=e^{-\frac{(1-\mu^2)\overline{d}_k^2}{2\sigma_r^2}}\left[\frac{\overline{d}_k \mu}{\sqrt{2\pi \sigma_r^2}}e^{-\frac{\overline{d}_k^2\mu^2}{2\sigma_r^2}}\right.\\
&\left.+\left(\frac{1}{2}+\frac{\overline{d}_k^2\mu^2}{2\sigma_r^2}\right)\left(1+{\rm erf}\left(\frac{\overline{d}_k \mu}{\sqrt{2}\sigma_r}\right)\right)\right].
\end{split}
\end{equation}
In the limit that $\sigma_r\gg d_k$, where the local-source dipole dominates over the kinematic dipole, $P(\mu|\sigma)$ approaches a uniform distribution: $P(\mu|\sigma)\rightarrow 1/2$. This is expected since the local-source dipole does not have a preferred direction (neglecting correlations with the kinematic dipole), and so the total dipole can, in this limit, point in any direction with equal probability.

On the other hand, if $\sigma_r\ll d_k$, the kinematic term dominates over the local-source contributions and $P(\mu|\sigma_r)$ exponentially increases with $\mu$, implying that the total measured dipole $\mathbf{d}$ should closely align with the direction of the kinematic dipole.

\section{Comparison with Previous Work}\label{A:NT15_comparison}
In this work, we employ a data-driven approach to directly infer $b(z)\frac{dN}{dz}$ for the NVSS sample using \textit{Tomographer} and cross-matching. This approach differs from previous studies that adopt separate parametric functional forms for $b(z)$ and $\frac{dN}{dz}$. The parametric approaches use fits to the auto-power measurements from the NVSS catalogs, along with spectroscopic redshift determinations from small radio galaxy surveys, and/or cosmological simulations \citep{2015ApJ...812...85N,2018JCAP...04..031B}. The purpose of this Appendix is to present explicit comparisons with some of the earlier studies.

\begin{figure}[ht!]
\begin{center}
\includegraphics[width=\linewidth]{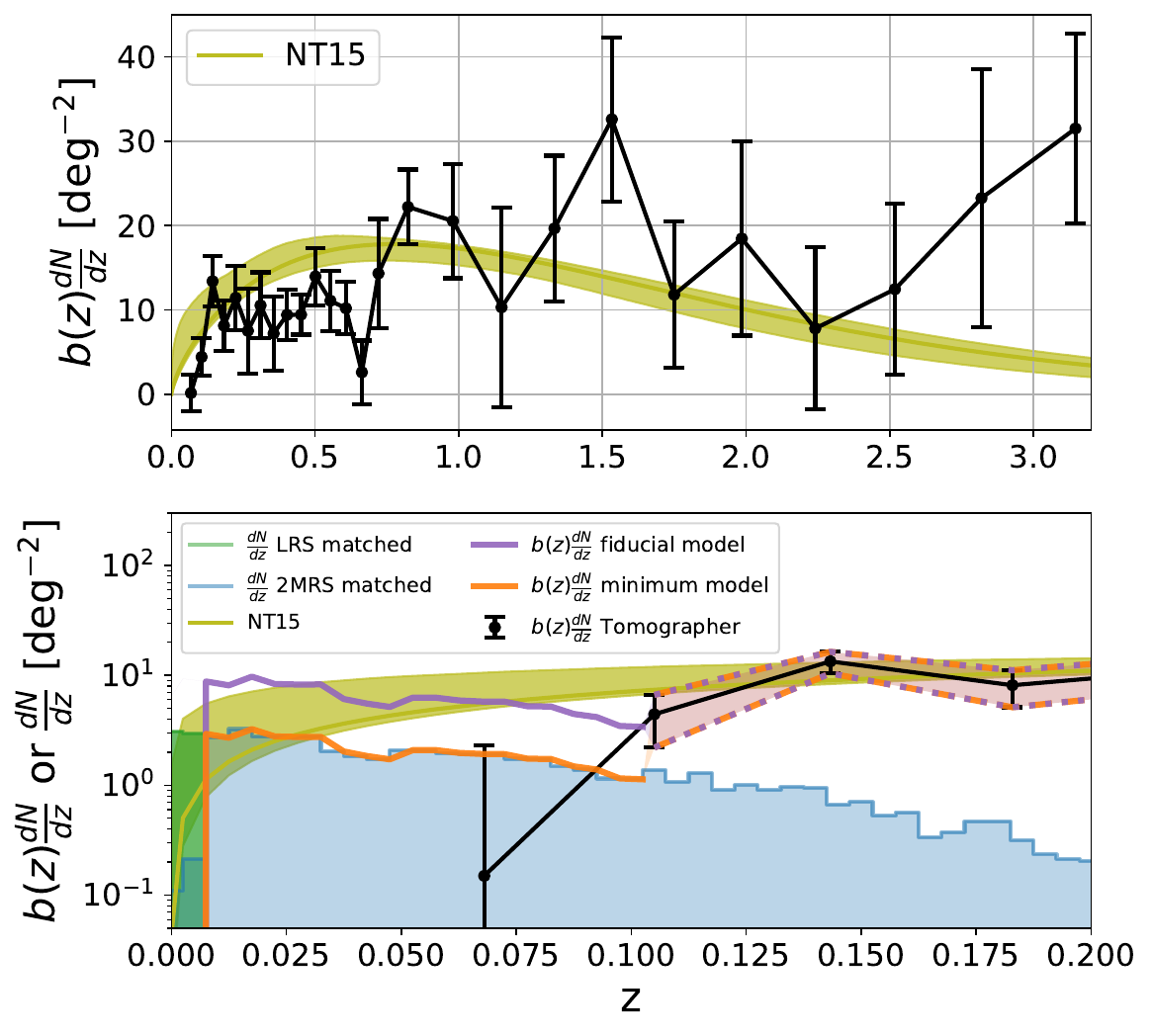}
\caption{\label{F:dNdz_NT15}
A comparison between our $b(z) \frac{dN}{dz}$ estimates and the model from NT15 (yellow). The solid yellow lines show the maximum likelihood constraints from that work, while the shaded regions give the $1\sigma$ allowed range. This band incorporates only the statistical errors in their parametric fits to $\frac{dN}{dz}$ and ignores further uncertainties in $b(z)$ and in the underlying parameteric description. The remaining lines and points are identical to those in Fig.~\ref{F:dNdz}. Note that the difference between the clustering redshift determinations and the NT15 parametric model at $z \gtrsim 1.5$ has a negligible impact on the estimates of the clustering dipole.}
\end{center}
\end{figure}

Specifically, we compare our results with the $b(z)\frac{dN}{dz}$ model of NT15 in Fig.~\ref{F:dNdz_NT15}. This model fits for $b(z)$ and $\frac{dN}{dz}$ using the redshift distributions for a small subset of NVSS radio galaixes with spectroscopic redshift determinations from the CENSORS and Hercules surveys, along with measurements of the auto-power spectrum of the NVSS sources. This model is subsequently used for studying the NVSS dipole in \citet{2016JCAP...03..062T}. We find broad overall consistency between the NT15 redshift distribution and our own determinations in Fig.~\ref{F:dNdz_NT15}. This is the case even though our estimate of the error band on the NT15 model is smaller than the true uncertainty on this model: for simplicity we adopt only the mean polynomial fit to $b(z)$ from \citet{2016JCAP...03..062T} and account for only the covariance in their $\frac{dN}{dz}$ parameters. Therefore, the error band on the NT15 model in Fig.~\ref{F:dNdz_NT15} underestimates the true uncertainty on the NT15 model. We further note that although the spectroscopic radio galaxies from CENSORS and Hercules provide direct, accurate redshift information, the sample is relatively small containing only 133 sources, which may not be fully representative of the NVSS populations. In contrast, our $z<0.1$ cross match with 2MRS contains $\sim 5000$ sources, and the \textit{Tomographer} estimates at higher redshift leverage the information available from large external spectroscopic tracer surveys, with for example 2.4 millions of sources from SDSS. Our more model-agnostic approach therefore complements the NT15 model and allows departures from the parametric forms assumed in the earlier work.

\begin{figure}[ht!]
\begin{center}
\includegraphics[width=\linewidth]{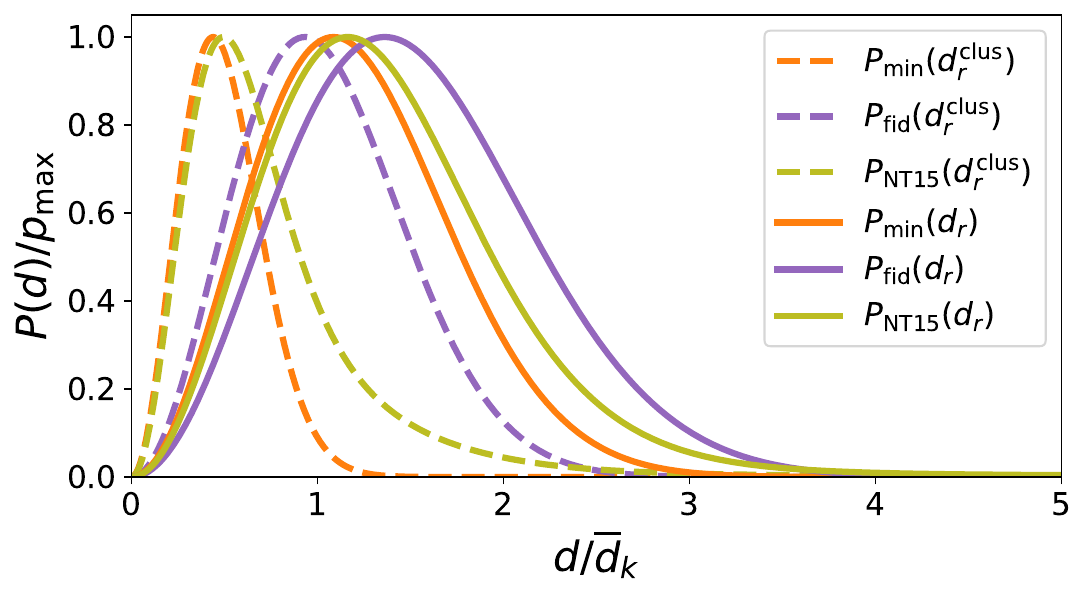}
\caption{\label{F:Pd_NT15} A comparison of the clustering/total dipole PDFs between our results and the NT15 $b(z) \frac{dN}{dz}$ model. 
The dashed lines show the clustering dipole PDF $P(d_r^{\rm clus})$, while the solid lines include the kinematic, clustering, and shot-noise contributions. The orange, purple, and yellow lines show the minimum, fiducial, and NT15 $b(z) \frac{dN}{dz}$ models, respectively. The different cases are fairly consistent, despite the different methodology employed in NT15 for estimating $b(z) \frac{dN}{dz}$.}
\end{center}
\end{figure}

Furthermore, Fig.~\ref{F:Pd_NT15} shows the resulting PDFs of the clustering and total dipole amplitudes, comparing our fiducial/minimum cases with calculations using $b(z) \frac{dN}{dz}$ from NT15. The total dipole PDF from the NT15 redshift distribution model exhibits a similar width to our two models, and the peak in the PDF lies in between our minimum/fiducial scenarios. This is the case even though the uncertainties adopted on $b(z) \frac{dN}{dz}$ are narrower at high redshift for the NT15 model. This occurs because the clustering dipole is mostly from local sources (at $z\lesssim0.2$), and more importantly, the width of the PDF is set mainly by sample/cosmic-variance in the clustering plus shot-noise amplitudes and by the range of possible alignment angles for the kinematic, clustering, and shot-noise dipole vector contributions.

As noted in Sec.~\ref{S:clus_dipole}, although the dipole PDF from the NT15 model is close to our minimum and fiducial models, it is unclear why \citet{2016JCAP...03..062T} -- which uses the NT15 clustering model -- report poorer agreement between their own measurements and models
than found here. Specifically, those authors use mock catalogs to account for the kinematic, clustering, and shot-noise contributions in assessing the compatibility of their dipole measurements with model expectations. They find a $2.1\sigma$ difference (for a flux cut of 15 mJy) between their measurement and model expectations, while we find 0.4-0.5$\sigma$ consistency for the same measurement and our own minimum and fiducial models (Table~\ref{T:dipole}), and also 0.4$\sigma$ consistency upon using the clustering dipole model from NT15.

It is unclear why this previous work finds less agreement than found here. As mentioned earlier (see Appendix~\ref{A:formalism} for details), the first two moments of our PDF agree with the expressions in \citet{2021JCAP...11..009N} (in the limit where we ignore our small correction for the width of the kinematic dipole distribution). 
A further cross-check of our significance level is hence that it approximately agrees with the $\sigma$-level expected from the first two moments in that work. Specifically, accounting for the full PDF in determining confidence intervals, we find 0.41/0.46$\sigma$ with our minimum/fiducial model (Table~\ref{T:dipole}), while the approximate $\sigma$-level estimate gives $0.66\sigma$. This agreement -- although inexact because the $\sigma$-level is a less accurate confidence interval estimate than that obtained from using the full PDF -- strengthens our confidence in our statistical significance determinations.

\section{Kinematic-clustering Dipole Correlation}\label{A:kin_clus_corr}
\subsection{Correlation Coefficient}
Here, we present the derivation of the $\langle\mathbf{v}_R\cdot\mathbf{d}_{\rm clus}\rangle$ and $\langle\mathbf{v}_R\cdot\mathbf{v}_R\rangle$ terms used in calculating the correlation coefficient between the kinematic and clustering dipoles (Eq.~\ref{E:vd_corr_coeff}).

By symmetry, we have $\langle\mathbf{v}_R\cdot\mathbf{d}_{\rm clus}\rangle=3\langle v_{R,z}d_{{\rm clus},z}\rangle$, and $\langle\mathbf{v}_R\cdot\mathbf{v}_R\rangle=3\langle v_{R,z}v_{R,z}\rangle$. Therefore, we only need to consider the $\mathbf{\hat{z}}$ direction in computing these inner products. From Eq.~\ref{E:vR_3D}, we obtain 
\begin{equation}
v_{R,z} = -if_\Omega H_0\int \frac{d^3\mathbf{k}}{(2\pi)^3}\frac{\mu_k}{k}W_v(kR)\delta_m^{\rm 3D}(\mathbf{k}),
\end{equation}
where $\mu_k\equiv k_z/k$. For the clustering dipole in the $\mathbf{\hat{z}}$ direction, $d_{{\rm clus},z}$, we expand the projected galaxy density field, $\delta_g^{\rm 2D}(\mathbf{\hat{x}})$, in spherical harmonics,
\begin{equation}
\delta_g^{\rm 2D}(\mathbf{\hat{x}}) = \sum_{\ell=0}^{\infty}\sum_{m=-\ell}^{m=\ell}a_{\ell m}Y_\ell^m(\mathbf{\hat{x}}),
\end{equation}
where $\mathbf{\hat{x}}$ denotes the direction unit vector. Since $Y_1^0(\hat{x})=\sqrt{\frac{3}{4\pi}}{\rm cos}\, \theta$, we have $d_{{\rm clus},z}=\sqrt{\frac{4\pi}{3}}a_{10}$. The 2D projected galaxy density field is obtained by integrating the 3D field with a window function $W(\chi)=\frac{1}{N}\frac{dN}{d\chi}$:
\begin{equation}
\begin{split}
\delta_g^{\rm 2D}(\mathbf{\hat{x}}) &= \int d\chi W(\chi)\delta_g^{\rm 3D}(\mathbf{x})\\
&= \int d\chi W(\chi)\int\frac{d^3\mathbf{k}}{(2\pi)^3}\delta_g^{\rm 3D}(\mathbf{k})e^{i\mathbf{k}\cdot\mathbf{x}},
\end{split}
\end{equation}
where $\mathbf{x}$ represents the 3D comoving coordinate, of magnitude $|\mathbf{x}|=\chi$, and its direction is defined by the unit vector $\mathbf{\hat{x}}$. Using the formula for the plane wave expansion in the spherical harmonic basis:
\begin{equation}
e^{i\mathbf{k}\cdot\mathbf{x}}=\sum_{\ell=0}^{\infty}\sum_{m=-\ell}^{m=\ell}4\pi i^\ell j_\ell(k\chi)Y_\ell^{m*}(\mathbf{\hat{k}})Y_\ell^m(\mathbf{\hat{x}}),
\end{equation}
we get
\begin{equation}\label{E:d_clusz}
d_{{\rm clus},z} = \sqrt{\frac{3}{4\pi}}4\pi i\int d \chi W(\chi)\int\frac{d^3\mathbf{k}}{(2\pi)^3}\delta_g^{\rm 3D}(\mathbf{k})j_\ell(k\chi)Y_1^{0*}(\mathbf{\hat{k}}),
\end{equation}
where $k$ and $\mathbf{\hat{k}}$ are the amplitude and direction unit vector of $\mathbf{k}$, respectively. With this, we can derive
\begin{equation}
\begin{split}
&\langle\mathbf{v}_R\cdot\mathbf{d}_{{\rm clus}}\rangle=3\langle v_{R,z}d_{{\rm clus},z}\rangle \\
&= 3\sqrt{\frac{3}{4\pi}}4\pi f_\Omega H_0\int d \chi W(\chi)\int\frac{d^3\mathbf{k}}{(2\pi)^3}j_\ell(k\chi)Y_\ell^{m*}(\mathbf{\hat{k}})\\
&\cdot\int \frac{d^3\mathbf{k'}}{(2\pi)^3}\frac{\mu_{k'}}{k'}W_v(k'R)\langle\delta_g^{\rm 3D}(\mathbf{k})\delta_m^{\rm 3D}(\mathbf{k'})\rangle.
\end{split}
\end{equation}
On large (linear) scales, yet neglecting GR corrections,
\begin{equation}
\langle\delta_g^{\rm 3D}(\mathbf{k})\delta_m^{\rm 3D}(\mathbf{k'})\rangle=(2\pi)^3\delta^D(\mathbf{k}+\mathbf{k'})b(z)D(z)P(k),
\end{equation}
where $b(z)$ is the bias factor, $D(z)$ is the linear growth factor, and $P(k)$ is the linear matter power spectrum at $z=0$. Using the relation $Y_1^{0*}(\mathbf{\hat{k}})=\sqrt{\frac{3}{4\pi}}\mu_k$,  and noting that we can integrate over $\mu_k$ by transforming to spherical coordinates, $d^3\mathbf{k}=2\pi k^2dkd\mu_k$, we obtain:
\begin{equation}\label{E:vR_dclus_corr}
\begin{split}
\langle\mathbf{v}_R\cdot\mathbf{d}_{\rm clus}\rangle=&3f_\Omega H_0\int\frac{dk}{2\pi^2}kW_v(kR)P(k)\\
&\cdot\int dz b(z)\frac{1}{N}\frac{dN}{dz}D(z)j_\ell[k\chi(z)].
\end{split}
\end{equation}

Similarly, for $\langle\mathbf{v}_R\cdot\mathbf{v}_R\rangle$, we get
\begin{equation}\label{E:vR_vR_corr}
\begin{split}
\langle\mathbf{v}_R\cdot\mathbf{v}_R\rangle&=-3(f_\Omega H_0)^2\int\frac{d^3\mathbf{k}}{(2\pi)^3}\frac{\mu_k}{k}W_v(kR)\\
&\int\frac{d^3\mathbf{k'}}{(2\pi)^3}\frac{\mu_{k'}}{k'}W_v(k'R)\langle\delta_m^{\rm 3D}(\mathbf{k})\delta_m^{\rm 3D}(\mathbf{k'})\rangle\\
&=(f_\Omega H_0)^2\int\frac{dk}{2\pi^2}W_v^2(kR)P(k),
\end{split}
\end{equation}
where we use the relation $\langle\delta_m^{\rm 3D}(\mathbf{k})\delta_m^{\rm 3D}(\mathbf{k'})\rangle=(2\pi)^3\delta^D(\mathbf{k}+\mathbf{k'})P(k)$ at $z=0$.

We also verified that using our expression for $d_{{\rm clus},z}$ from Eq.~\ref{E:d_clusz}, we recover the common expression,
\begin{equation}
\langle\mathbf{d}_{\rm clus}\cdot\mathbf{d}_{\rm clus}\rangle=3 \, \langle d_{{\rm clus},z}d_{{\rm clus},z}\rangle = \frac{9}{4\pi}C_1^{\rm clus},
\end{equation}
with $C_1^{\rm clus}$ following Eq.~\ref{E:Clclus}.

\subsection{PDF with Kinematic-clustering Dipole Correlation}
\label{S:pdf_kin_clus_included}
\begin{figure}[ht!]
\begin{center}
\includegraphics[width=\linewidth]{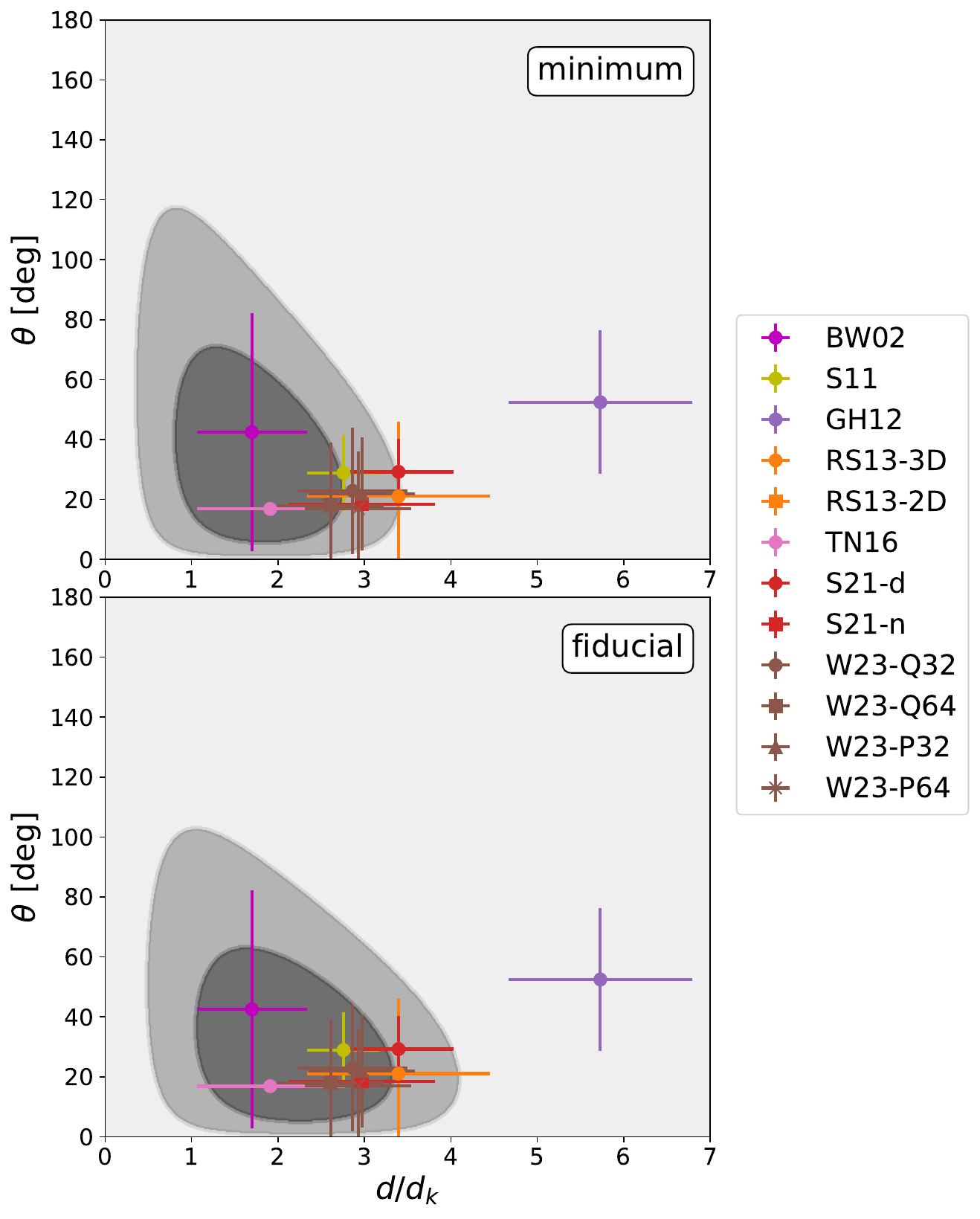}
\caption{\label{F:Pdth_vdcorr} This  figure is identical to Fig.~\ref{F:Pdth} except here the model PDFs include the kinematic-clustering correlation which shifts the model PDFs toward higher dipole amplitudes and smaller offset angles.}
\end{center}
\end{figure}

\begin{figure}[ht!]
\begin{center}
\includegraphics[width=\linewidth]{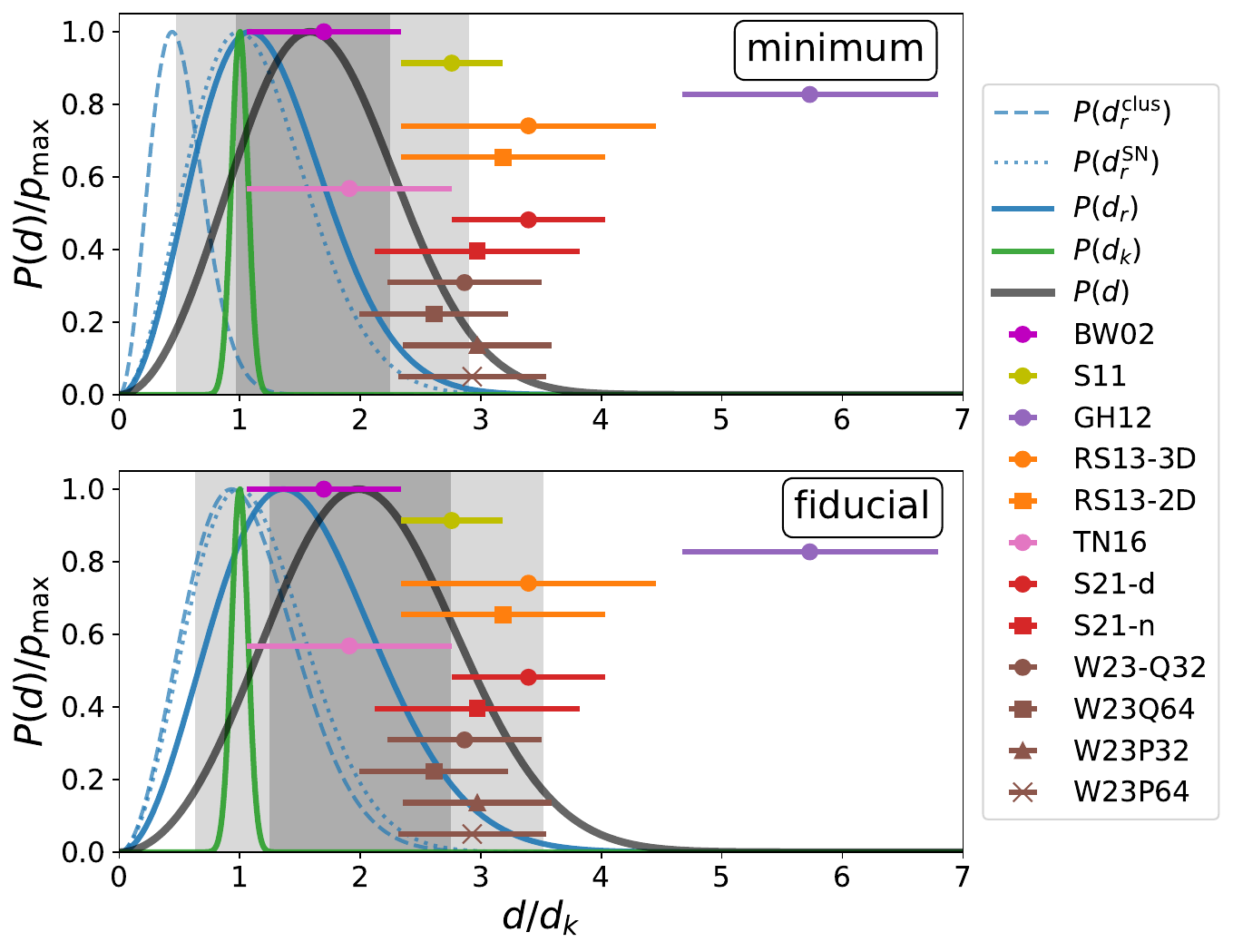}
\caption{\label{F:Pd_vdcorr} This figure is identical to Fig.~\ref{F:Pd} except here the model PDFs include the kinematic-clustering correlation which shifts the 1D PDFs toward higher dipole amplitudes.} 
\end{center}
\end{figure}

\begin{figure}[ht!]
\begin{center}
\includegraphics[width=\linewidth]{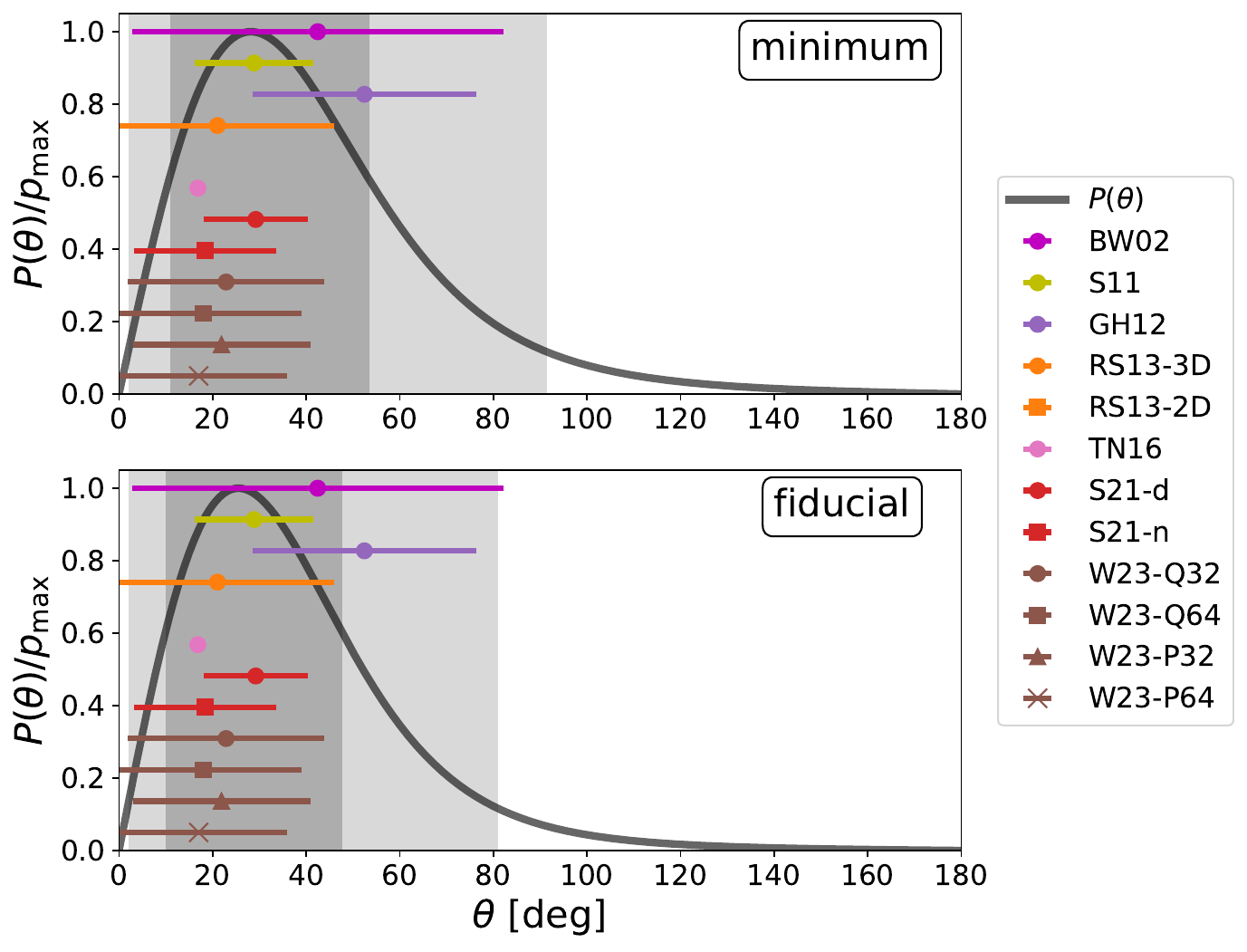}
\caption{\label{F:Pth_vdcorr} This figure is identical to Fig.~\ref{F:Pth} except here the model PDFs include the kinematic-clustering correlation which shifts the 1D PDFs toward smaller offset angles.} 
\end{center}
\end{figure}

Given the two dipole vectors, $\mathbf{d}_r$ and $\mathbf{d}_k$, and the correlation coefficient, $r_{\rm vd}$, the conditional distribution of $\mathbf{d}_r$ given $\mathbf{d}_k$ in Eq.~\ref{E:P(d_r)_3D} needs to be modified to
\begin{equation}\label{E:P(d_r|d_k)_3D}
P(\mathbf{d}_r|\mathbf{d}_k,\sigma_r) = \frac{1}{[2\pi\sigma_r^2(1-r_{\rm vd}^2)]^{3/2}}e^{-\frac{d_{r,x}^2+d_{r,y}^2+(d_{r,z}-d_{\rm vd})^2}{2\sigma_r^2(1-r_{\rm vd}^2)}},
\end{equation}
where
\begin{equation}
d_{\rm vd} = r_{\rm vd}\frac{\sigma_r}{\sigma_{\rm vd}}d_k,
\end{equation}
and
\begin{equation}
\sigma_{\rm vd}^2=\frac{1}{3  }A^2\frac{<\mathbf{v}\cdot\mathbf{v}>}{c^2},
\end{equation}
which is the kinematic dipole variance corresponding to the calculated velocity variance, and we take $A=2+\overline{x}(1+\overline{\alpha})=3.82$ (Sec.~\ref{S:kinematic_dipole}). With this, the PDF of the total dipole vector becomes (Eq.~\ref{E:P(d_3D)})
\begin{equation}\label{E:P(d_3D)_vdcorr}
\begin{split}
P(\mathbf{d}|\sigma_r)=\frac{1}{2\pi\Tilde{\sigma}_r^2\sqrt{2\pi(\Tilde{\sigma}_r^2+\sigma_k^2)}}e^{-\frac{d_x^2+d_y^2}{2\Tilde{\sigma}_r^2}}e^{-\frac{(d_z-\Tilde{d}_k)^2}{2(\Tilde{\sigma}_r^2+\sigma_k^2)}},
\end{split}
\end{equation}
where $\Tilde{\sigma}_r^2\equiv\sigma_r^2(1-r_{\rm vd}^2)$ and $\Tilde{d}_k\equiv\overline{d}_k + d_{\rm vd}$. Note that this expression has the same form as Eq.~\ref{E:P(d_3D)} with the replacements $\sigma_r^2\rightarrow\Tilde{\sigma}_r^2$ and $\overline{d}_k\rightarrow\Tilde{d}_k$. Therefore, $P(d,\theta)$, $P(d)$, and $P(\theta)$ also follow the same expression as in Sec.~\ref{A:formalism} with these replacements for $\sigma_r^2$ and $\overline{d}_k$. The resulting PDFs after taking into account the kinematic-clustering correlation are shown in Fig.~\ref{F:Pdth_vdcorr}, \ref{F:Pd_vdcorr}, and \ref{F:Pth_vdcorr}. As detailed in Table~\ref{T:dipole}, this
further reduces the difference with the NVSS measurements, although the shifts are generally fairly small. 

\section{Bulk Flow Velocity}\label{A:bulk_flow}
The rms of the bulk velocity, averaged over the NVSS sources, is given by the square of the weighted sum of velocity vectors,
\begin{equation}
\langle\mathbf{v}^2_{\rm bulk}\rangle=\langle\left(\int d^3\mathbf{x}W^{\rm 3D}(\mathbf{x})\mathbf{v}(\mathbf{x})\right)^2\rangle,
\end{equation}
where $\mathbf{x}$ is the 3D comoving coordinate, and $W^{\rm 3D}(\mathbf{x})$ is the window function defined in the 3D space. 

The 3D window function, $W^{\rm 3D}(\mathbf{x})$, is normalized such that $\int d^3\mathbf{x} \, W^{\rm 3D}(\mathbf{x}) = 1$. Assuming the window function depends only on the line-of-sight comoving distance, $\chi$, we can convert the integration to spherical coordinates. The normalization condition then becomes $\int d\chi \, 4\pi\chi^2 \, W^{\rm 3D}(\chi) = 1$. This allows us to relate the 3D window function to the commonly defined 1D window function, $W(\chi) = 4\pi\chi^2 W^{\rm 3D}(\chi)$, which is normalized to unity when integrated over the line-of-sight comoving distance: $\int d\chi \, W(\chi) = 1$. For a source count field, as in our case, the 1D window function is given by $W(\chi) = \frac{1}{N}\frac{dN}{d\chi}$.

By symmetry, the expected rms in the $\mathbf{\hat{x}}$, $\mathbf{\hat{y}}$, and $\mathbf{\hat{z}}$ directions are equal. Therefore, the rms of the bulk velocity can be computed as follows:
\begin{equation}
\begin{split}
\langle\mathbf{v}^2_{\rm bulk}\rangle&=3\langle\left(\int d^3\mathbf{x}W^{\rm 3D}(\mathbf{x})v_z(\mathbf{x})\right)^2\rangle\\
&=3\int d^3\mathbf{x} W^{\rm 3D}(\mathbf{x})\int d^3\mathbf{x}'W^{\rm 3D}(\mathbf{x}')\langle v_z(\mathbf{x})v_z(\mathbf{x}')\rangle\\
&=3\int d^3\mathbf{x} W^{\rm 3D}(\mathbf{x})\int d^3\mathbf{x}'W^{\rm 3D}(\mathbf{x}')\\
&\cdot\int\frac{d^3\mathbf{k}}{(2\pi)^3}e^{i\mathbf{k}\cdot\mathbf{x}}\int\frac{d^3\mathbf{k}'}{(2\pi)^3}e^{i\mathbf{k}'\cdot\mathbf{x}'}\langle v_z(\mathbf{k})v_z(\mathbf{k}')\rangle.
\end{split}
\end{equation}
From the continuity equation and linear perturbation theory, we get
\begin{equation}
v_z(\mathbf{k})=-\frac{\mu}{k}\frac{iH(z)f_\Omega(z)}{1+z}\delta_m(\mathbf{k}),
\end{equation}
where $\mu=k_z/k$, $H(z)$ is the Hubble parameter, and $f_\Omega(z)$ is the growth rate. Therefore, we obtain
\begin{equation}
\begin{split}
\langle v_z(\mathbf{k})v_z(\mathbf{k}')\rangle=&\frac{\mu^2}{k^2}\left(\frac{H(z)f_\Omega(z)D(z)}{1+z}\right)\left(\frac{H(z')f_\Omega(z')D(z')}{1+z'}\right)\\
&\cdot(2\pi)^3\delta^D(\mathbf{k}+\mathbf{k'})P(k),
\end{split}
\end{equation}
where $D(z)$ is the linear growth factor, and $P(k)$ is the linear theory matter power spectrum at $z=0$. With this, we derive
\begin{equation}
\begin{split}
&\langle\mathbf{v}^2_{\rm bulk}\rangle
=3\int d^3\mathbf{x} W^{\rm 3D}(\mathbf{x})\int d^3\mathbf{x}'W^{\rm 3D}(\mathbf{x}')\int\frac{d^3\mathbf{k}}{(2\pi)^3}\frac{\mu^2}{k^2}e^{i\mathbf{k}\cdot(\mathbf{x}-\mathbf{x}')}\\
&\cdot P(k)\left(\frac{H(z)f_\Omega(z)D(z)}{1+z}\right)\left(\frac{H(z')f_\Omega(z')D(z')}{1+z'}\right)\\
&=3\int\frac{d^3\mathbf{k}}{(2\pi)^3}\frac{\mu^2}{k^2}P(k)\\
&\cdot\left|\int d^3\mathbf{x} W^{\rm 3D}(\mathbf{x})\left(\frac{H(z)f_\Omega(z)D(z)}{1+z}\right)e^{i\mathbf{k}\cdot\mathbf{x}}\right|^2.
\end{split}
\end{equation}
Using the relation between the 3D and 1D window functions, and evaluating the last integral in spherical coordinates, we arrive at the following expression:
\begin{equation}
\begin{split}
&\int d^3\mathbf{x} W^{\rm 3D}(\mathbf{x})\left(\frac{H(z)f_\Omega(z)D(z)}{1+z}\right)e^{\pm i\mathbf{k}\cdot\mathbf{x}} \\
=&\int d\chi d\mu\, 2\pi\chi^2 W^{\rm 3D}(\chi)\left(\frac{H(z)f_\Omega(z)D(z)}{1+z}\right)e^{\pm ik\chi\mu}\\
=&\int d\chi\, 2\pi\chi^2 \frac{W^{\rm 1D}(\chi)}{4\pi\chi^2}\left(\frac{H(z)f_\Omega(z)D(z)}{1+z}\right)\int_{-1}^{1} d\mu e^{\pm ik\chi\mu}\\
=&\int d\chi\, 2\pi\chi^2 \frac{W^{\rm 1D}(\chi)}{4\pi\chi^2}\left(\frac{H(z)f_\Omega(z)D(z)}{1+z}\right)\frac{2{\rm sin}(k\chi)}{k\chi},\\
=& \int dz \frac{1}{N}\frac{dN}{dz}\left(\frac{H(z)f_\Omega(z)D(z)}{1+z}\right)\frac{{\rm sin}[k\chi(z)]}{k\chi(z)}.
\end{split}
\end{equation}
Similarly, by converting the 3D integration in $k$ into spherical coordinates and integrating over $\mu$, we obtain the following expression for the bulk velocity variance:\footnote{We note that \citet{2021JCAP...11..009N} also discussed the bulk flow velocity and presented the same formalism in their Eq. D.2. However, there is a missing factor of $1+z$ in their denominator.}:
\begin{equation}\label{E:v_bulk}
\begin{split}
\langle\mathbf{v}^2_{\rm bulk}\rangle&=\frac{1}{2\pi^2}\int dk P(k)\\
&\cdot\left[\int dz \frac{1}{N}\frac{dN}{dz}\left(\frac{H(z)f_\Omega(z)D(z)}{1+z}\right)\frac{{\rm sin}[k\chi(z)]}{k\chi(z)}\right]^2.
\end{split}
\end{equation}

\bibliography{dipole}{}
\bibliographystyle{aasjournal}

\end{document}